\documentclass[aps,eqsecnum,superscriptaddress]{revtex4}
\usepackage{epsfig,rotating,amsmath,amssymb,graphics,rotate,psfrag}
\usepackage{minitoc}
\pdfoutput=1

\newcommand{\cambridge}{Astrophysics Group, Cavendish Laboratory, J J
  Thomson Avenue, Cambridge, CB3 0HE UK} 
\newcommand{\ipac}{Spitzer Science Center, MS 220-6, California
  Institute of Technology, Pasadena, CA 91125 USA}
\newcommand{\manchester}{Alan Turing Building, School of
  Physics and Astronomy, Oxford Road, Manchester M13 9PL, UK}
\newcommand{\bristol}{HH Wills Physics Laboratory, University of
  Bristol, Tyndall Avenue, Bristol BS8 1TL, UK}
\newcommand{\leicester}{Department of Physics and Astronomy,
  University of Leicester, LE1 7RH Leicester, UK} 
\newcommand{\ucl}{Optical Science Laboratory, University College
  London, Gower Street, London, UK} 
\newcommand{\uclast}{Department of Physics \& Astronomy, University
  College London, Gower Street, London WC1E 6BT, UK}
\newcommand{\durham}{Department of Physics, University of Durham,
  South Road, Durham DH1 3LE, UK} 
\newcommand{\kavli}{Kavli Institute for Cosmology Cambridge, Madingley
  Road, Cambridge CB3 0HA, UK} 
\newcommand{\southampton}{Physics and Astronomy, University of
  Southampton, Southampton SO17 1BJ, UK} 
\newcommand{\herts}{Centre for Astrophysics Research, Science \&
  Technology Research Institute, University of Hertfordshire,
  Hatfield, UK}
\newcommand{\oxford}{Astrophysics, Department of Physics, University
  of Oxford, Keble Road, Oxford OX1 3RH, UK}
\newcommand{\openuniversity}{Department of Physics and Astronomy, The
  Open University, Walton Hall, Milton Keynes, MK7 6AA, UK}
\newcommand{\ral}{RAL Space, STFC Rutherford Appleton Laboratory,
  Chilton, Didcot, Oxfordshire, OX11 0QX, UK}
\newcommand{\birmingham}{School of Physics and Astronomy, University
  of Birmingham, Edgbaston, Birmingham, B15 2TT, UK} 
\newcommand{\erlangen}
{Dr Karl-Remeis-Sternwarte and Erlangen Center for Astroparticle
  Physics, Sternwartstr. 7, 96049 Bamberg, Germany} 
\newcommand{\copernicus}{Nicolaus Copernicus Astronomical Center,
  ul. Bartycka 18, 00-716 Warsaw, Poland}
\newcommand{\malta}{Department of Physics, Faculty of Science,
  University of Malta, Malta} 
\newcommand{\cape}{Physics Department, University of the
  Western Cape, Bellville 7535, South Africa}
\newcommand{\capetown}{Astrophysics, Cosmology and Gravitation Centre,
  University of Cape Town, Rondebosch 7701, Cape Town, South Africa}
\newcommand{\curtin}{International Centre for Radio Astronomy Research
  - Curtin University, GPO Box U1987, Perth, WA 6845, Australia}
\newcommand{\jive}{Joint Institute for VLBI in Europe (JIVE), Postbus
  2, 7990 AA Dwingeloo, The Netherlands}
\newcommand{\caltech}{Cahill Center for Astronomy and Astrophysics,
  California Institute of Technology, Pasadena, CA 91125, USA}
\newcommand{\goddard}{CRESST and NASA Goddard Space Flight Center,
  Code 661, Greenbelt, MD 20771,USA} 
\newcommand{\cea}{Laboratoire AIM, CEA/IRFU - Universite
  Paris Diderot, CEA DSM/IRFU/SAp, F-91191 Gif-sur-Yvette, France}
\newcommand{\esochile}{European Southern Observatory, ESO Vitacura, Alonso
  de Cordova 3107, Vitacura, Casilla 19001, Santiago, Chile}
\newcommand{\esogermany}{European Southern Observatory Karl-Schwarzschild-Str
  2, 85748 Garching, Germany}
\newcommand{\aalto}{ Aalto University Mets\"{a}hovi Radio
  Observatory, Metsahovintie 114, FIN-02540 Kylmala, Finland}
\newcommand{\csiro}{CSIRO Astronomy \& Space Science, Australia
  Telescope National Facility, PO Box 76, Epping, NSW 1710, Australia}
\newcommand{\inaf}{INAF/IASF-Roma, I-00133 Roma, Italy}
\newcommand{\berkeley}{Space Sciences Laboratory, 7 Gauss Way,
  University of California, Berkeley, CA 94720-7450, USA}
\newcommand{\amsterdam}{Astronomical Institute Anton Pannekoek,
  University of Amsterdam, Science Park 904, 1098XH Amsterdam, The Netherlands}
\newcommand{\sydney}{Sydney Institute for Astronomy, School of Physics
  A28, University of Sydney, NSW 2006, Australia}  
\newcommand{\iaa}{Instituto de Astrof\'{i}sica de Andalucia (IAA-CSIC),
Glorieta de la Astronomia s/n, 18.008 Granada, Spain}
\newcommand{\iac}{Instituto de Astrof\'{i}sica de Canarias,V\'{i}a
  L\'{a}ctea, 38205 La Laguna, Tenerife, Spain}
\newcommand{\alma}{ALMA Santiago Central Offices,
  Alonso de Cordova 3107, Vitacura, Casilla 7630355, Santiago, Chile}
\newcommand{\macarthur}{Division of Physics, Mathematics and
  Astronomy, California Institute of Technology, Pasadena, CA 91125, USA}
\newcommand{\hoofddocent}{Astronomical Institute Anton
  Pannekoeka, University of Amsterdam, Postbus 94249, 1090 GE
  Amsterdam, the Netherlands} 
\newcommand{\smithsonian}{Smithsonian Astrophysical Observatory, 60
  Garden Street, Cambridge, MA 02138-1516, USA}
\newcommand{\michigan}{Department of Astronomy, University of
  Michigan, 500 Church St, Ann Arbor, MI 48109-1042, USA}
\newcommand{\bucharest}{Institute for Space Sciences, Bucharest,
  Romania}
\newcommand{\dublin}{Dublin City University, Dublin 9, Ireland; Dublin
  Institute for Advanced Studies, 31 Fitzwilliam Place, Dublin 2,
  Ireland} 
\newcommand{\florida}{Department of Physics and Space Sciences,
  Florida Institute of Technology, 150 W. University Blvd., Melbourne,
  FL, USA} 
 
\newcommand{\barcelona}{Departament d'Astronomia i Meteorologia,
  Universitat de Barcelona, Martíi Franquès 1, 08028 Barcelona, Spain} 
\newcommand{\purdue}{Purdue University, West Lafayette, IN 47907, USA}
\newcommand{\exeter}{School of Physics, University of Exeter, Stocker
  Road, Exeter EX4 4QL} 

\begin{document}

\title{Future Science Prospects for AMI}


\author{Keith Grainge}
\affiliation{\cambridge}
\affiliation{\kavli}
\author{Paul Alexander}
\affiliation{\cambridge}
\affiliation{\kavli}
\author{Richard Battye}
\affiliation{\manchester}
\author{Mark Birkinshaw}
\affiliation{\bristol}
\author{Andrew Blain}
\affiliation{\leicester}
\author{Malcolm Bremer}
\affiliation{\bristol}
\author{Sarah Bridle}
\affiliation{\uclast}
\author{Michael Brown}
\affiliation{\manchester}
\author{Richard Davis}
\affiliation{\manchester}
\author{Clive Dickinson}
\affiliation{\manchester}
\author{Alastair Edge}
\affiliation{\durham}
\author{George Efstathiou}
\affiliation{\kavli}
\author{Robert Fender}
\affiliation{\southampton}
\author{Martin Hardcastle}
\affiliation{\herts}
\author{Jennifer Hatchell}
\affiliation{\exeter}
\author{Michael Hobson}
\affiliation{\cambridge}
\author{Matthew Jarvis}
\affiliation{\oxford}
\affiliation{\cape}
\author{Benjamin Maughan}
\affiliation{\bristol}
\author{Ian McHardy}
\affiliation{\southampton}
\author{Matthew Middleton}
\affiliation{\durham}
\author{Anthony Lasenby}
\affiliation{\cambridge}
\affiliation{\kavli}
\author{Richard Saunders}
\affiliation{\cambridge}
\affiliation{\kavli}
\author{Giorgio Savini}
\affiliation{\ucl}
\author{Anna Scaife}
\affiliation{\southampton}
\author{Graham Smith}
\affiliation{\birmingham}
\author{Mark Thompson}
\affiliation{\herts}
\author{Glenn White}
\affiliation{\ral}
\affiliation{\openuniversity}
\author{Kris Zarb-Adami}
\affiliation{\oxford}
\affiliation{\malta}

\author{James Allison}
\affiliation{\sydney}
\author{Jane Buckle}
\affiliation{\cambridge}
\author{Alberto Castro-Tirado}
\affiliation{\iaa}
\author{Maria Chernyakova}
\affiliation{\dublin}
\author{Roger Deane}
\affiliation{\capetown}
\author{Farhan Feroz}
\affiliation{\cambridge}
\author{Ricardo G\'{e}nova-Santos}
\affiliation{\iac}
\author{David Green}
\affiliation{\cambridge}
\author{Diana Hannikainen}
\affiliation{\florida}
\affiliation{\aalto}
\author{Ian Heywood}
\affiliation{\oxford}
\author{Natasha Hurley-Walker}
\affiliation{\curtin}
\author{R\"{u}diger Kneissl}
\affiliation{\esochile}
\affiliation{\alma}
\author{Karri Koljonen}
\affiliation{\aalto}
\author{Shrinivas Kulkarni}
\affiliation{\macarthur}
\author{Sera Markoff}
\affiliation{\hoofddocent}
\author{Carrie MacTavish}
\affiliation{\kavli}
\author{Michael McCollough}
\affiliation{\smithsonian}
\author{Simone Migliari}
\affiliation{\barcelona}
\author{Jon M. Miller}
\affiliation{\michigan}
\author{James Miller-Jones}
\affiliation{\curtin}
\author{Malak Olamaie}
\affiliation{\cambridge}
\author{Zsolt Paragi}
\affiliation{\jive}
\author{Timothy Pearson}
\affiliation{\caltech}
\author{Guy Pooley}
\affiliation{\cambridge}
\author{Katja Pottschmidt}
\affiliation{\goddard}
\author{Rafael Rebolo}
\affiliation{\iac}
\author{John Richer}
\affiliation{\cambridge}
\author{Julia Riley}
\affiliation{\cambridge}
\author{J\'{e}r\^{o}me Rodriguez}
\affiliation{\cea}
\author{Carmen Rodr\'{i}guez-Gonz\'{a}lvez}
\affiliation{\ipac}
\author{Anthony Rushton}
\affiliation{\esogermany}
\author{Petri Savolainen}
\affiliation{\aalto}
\author{Paul Scott}
\affiliation{\cambridge}
\author{Timothy Shimwell}
\affiliation{\csiro}
\author{Marco Tavani}
\affiliation{\inaf}
\author{John Tomsick}
\affiliation{\berkeley}
\author{Valeriu Tudose}
\affiliation{\bucharest}
\author{Kurt van der Heyden}
\affiliation{\capetown}
\author{Alexander van der Horst}
\affiliation{\amsterdam}
\author{Angelo Varlotta}
\affiliation{\purdue}
\author{Elizabeth Waldram}
\affiliation{\cambridge}
\author{Joern Wilms}
\affiliation{\erlangen}
\author{Andrzej Zdziarski}
\affiliation{\copernicus}
\author{Jonathan Zwart}
\affiliation{\cape}
\author{Yvette Perrott}
\affiliation{\cambridge}
\author{Clare Rumsey}
\affiliation{\cambridge}
\author{Michel Schammel}
\affiliation{\cambridge}

\normalsize

\begin{abstract}

\begin{center}
{\bf \large Abstract} \\
\end{center}

The Arcminute Microkelvin Imager~(AMI)~\citep{jon08} is a telescope
specifically designed for high sensitivity measurements of
low-surface-brightness features at cm-wavelength and has unique,
important capabilities. It consists of two interferometer arrays
operating over 13.5--18~GHz that image structures on scales of 0.5--10
arcmin with very low systematics.  The Small Array (AMI-SA; ten 3.7-m
antennas) couples very well to Sunyaev-Zel'dovich~(SZ) features from
galaxy clusters and to many Galactic features.  The Large
Array~(AMI-LA; eight 13-m antennas) has a collecting area ten times
that of the AMI-SA and longer baselines, crucially allowing the
removal of the effects of confusing radio point sources from regions
of low surface-brightness, extended emission. Moreover AMI provides
fast, deep object surveying and allows monitoring of large numbers of
objects. In this White Paper we review the new science --- both
Galactic and extragalactic --- already achieved with AMI and outline
the prospects for much more.

\end{abstract}

\maketitle

\begin{center}
{\bf \large Contents} \\
\end{center}

\begin{center}
\begin{tabular}{p{15mm}p{80mm}}
\hline
\ref{sz} &Sunyaev-Zel'dovich Effect from Galaxy Clusters\\
\ref{ami-planck} &AMI's role in \textit{Planck} Validation and
Follow-up\\
\ref{galactic} &Galactic research with AMI\\
\ref{transients} &Transient and Variable-Source Astrophysics\\
\ref{upgrades} &Upgrades to AMI\\
\ref{future} &Future Operational Model\\
\hline
\end{tabular}
\end{center}

\section{Sunyaev-Zel'dovich Effect from Galaxy Clusters}\label{sz}

\subsection{Introduction}

Clusters of galaxies are the most massive bound structures in the
cosmic web of large-scale structure.  They evolve through the linear
regime from the density perturbations imaged in the primordial cosmic
microwave background (CMB) and then continue to grow in the non-linear
regime through in-fall and mergers.  As such, clusters have great
potential for use in cosmology, both to investigate the aspects of the
underlying cosmological model (e.g. $\sigma_8$) and as cosmographic
buoys to study the evolution of structure as a whole in the universe
(see e.g. \citet{2011ARA&A..49..409A}).  In order to connect clusters
to the underlying cosmology, the critical parameter that must be
measured is their mass. Various scaling relations have been proposed
for mass determination, but the Comptonisation $Y$ parameter is now
recognised as a highly-reliable mass indicator
\citep{2006ApJ...650..128K}. This is, of course, measured directly
through SZ measurements, without the need for a model-dependent
calculation from the data as is the case for X-rays.  The
redshift-independence of SZ surface brightness offers the possibility
of observing clusters right back to their epoch of formation; and the
linear (as opposed to quadratic) dependence of the SZ signal on gas
density allows imaging of the outer regions of the Intra Cluster
Medium~(ICM) --- \citet{2011ARA&A..49..409A} emphasise the importance
of investigating these outer regions, which is challenging in X-rays,
but straightforward in SZ. Combining SZ measurements with X-ray
observations and optical-lensing data gives essential, new
information.

In addition to their use for cosmology, clusters are of fundamental
interest in their own right, harbouring a great deal of detailed ICM
astrophysics (see e.g. \citet{2011MNRAS.410.2446K}). For example,
merger events are laboratories in which the nature of dark matter can
be tested \citep{2006ApJ...648L.109C}.  Imaging of detailed ICM
structure is now becoming possible in both X-ray and SZ
(e.g. \citet{2010ApJ...716..739M,2011ApJ...734...10K}), allowing
measurement of density and temperature over a range of scales.  A
critical issue in understanding the ICM is the role of magnetic
field. This could substantially stiffen the ICM equation of state,
affect cluster dynamics and evolution, and lead to bias in many mass
estimations. Little is known about ICM magnetic field, because the
requisite measurements have been out of reach: however, the
instrumental situation is now changing.

The initial science drivers for AMI were, via Sunyaev-Zel'dovich
effect~(SZ) measurements, to constrain cosmology through conducting a
blind survey for galaxy clusters; and to explore the physics of
clusters and tie down their scaling relations and evolution through
pointed observations towards known or candidate clusters. AMI has made
significant advances in both of these areas.

\subsection{Blind cluster SZ surveys}

\begin{figure}[t]
\begin{center}
\includegraphics[width=11.0cm]{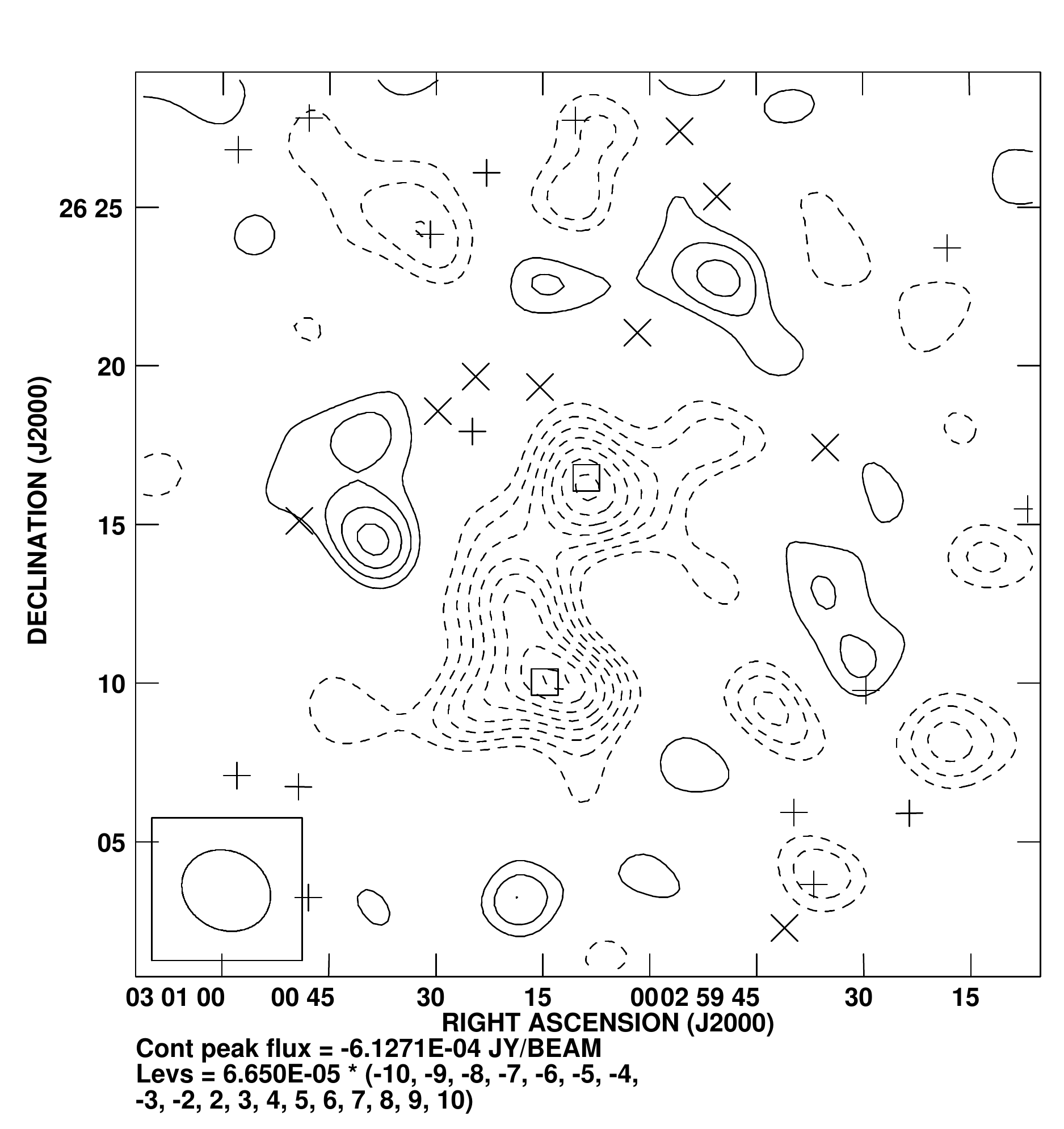}
\caption{Map of the SZ effect in the first cluster system to be
  discovered in the AMI blind survey. positive contours are solid and
  negative ones dashed; contour levels are in units of $\sigma$.
  Crosses (vertical and diagonal) indicate the position of point radio
  sources that have been subtracted from the data. Squares show the
  best fit positions for the clusters.  The synthesised beam is
  indicated in the lower-left corner.}
\label{firstblind}
\end{center}
\end{figure}

We have surveyed 10 first-look fields to target sensitivities of
$100\,\mu$Jy and $60\,\mu$Jy with AMI's Small (SA) and Large (LA)
arrays respectively. Each candidate cluster identified through this
analysis is then reobserved with deep, pointed observations to verify
the detection.  In \citet{tim10} we describe the first successful
cluster detection from the first (blind) field to be completed (see
Figure~\ref{firstblind}).  This distant, remarkably extended system,
probably a merger, is detected at very high significance by AMI in
SZ. We calculate that the cluster mass limit to which our survey will be
sensitive as $2\times10^{14} M_\odot$ by scaling from the reported
detection.  

A full Bayesian analysis \citep{2012MNRAS.421.1136A} of the first 10
fields is underway to detect clusters, determine their significance
and place constraints on cosmological parameters, e.g.  $\sigma_8$.
In addition, the AMI-LA observations of our fields have allowed us to
measure the radio source counts at 15.7~GHz down to 0.5~mJy
\citep{2011MNRAS.415.2699A,2011MNRAS.415.2708A}; this 10C survey is
the deepest radio survey of any significant extent above 1.4~GHz. New
programme of observations will push the source counts a further factor
of three deeper. Preliminary results suggest the detection of a change
in the slope of the source counts and a change in the spectral index
distribution, both of which indicate a change in the radio source
population.

AMI SZ-surveying and analysis is now routine and has been demonstrated
to detect clusters down to a limiting mass of $2\times10^{14}
M_\odot$. Therefore there is a great deal of scientific potential in
performing complementary surveys towards fields observed in other
wavebands.  For example, the AMI blind surveys have now been extended
to include additional fields toward the Lockmann Hole and ELAIS N1
regions which have been surveyed deeply by UKIDSS. This ongoing
project will allow cross-calibration of IR and SZ scaling
relationships and investigation of cluster selection functions of the
two methods.  This project will also provide a legacy survey of these
well-studied regions.  Another example of a future collaboration is
with Battye et al.~on the SuperCLASS legacy programme
(http://www.e-merlin.ac.uk/legacy/projects/superclass.html), which has
recently been allocated 862 hours of eMERLIN observing time. The aim
of this programme is to measure weak gravitation lensing through radio
imaging towards a supercluster field. Combining these data with SZ
measurements from AMI will set very powerful joint constraints on
parameters of baryonic and dark matter structures in this field.

\subsection{Pointed SZ observations}

AMI has been used for a wide range of SZ projects targeting fields of
interest revealed in other wavebands
\citep{ami06,tash11,carmen11a,carmen11b,jon11}.  We have
demonstrated the capabilities of AMI through observations of a variety
of different known clusters \citep{jon11}. We have also used these
observations to validate our fully Bayesian analysis tool, {\sc
  McAdam}~\citep{2009MNRAS.398.2049F}, which uses nested sampling to
fit various cluster parameterisations efficiently to our data, in
particular generating a robust estimate of the key observable of
cluster mass out to the virial radius.

\begin{figure}[t]
\begin{center}
\includegraphics[width=11.0cm]{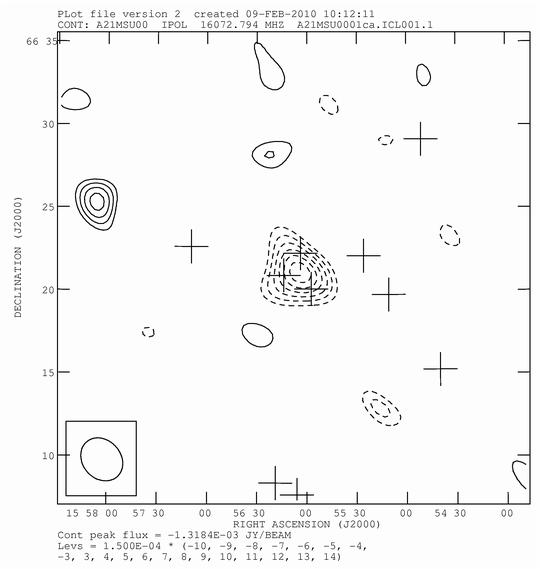}
\caption{AMI map of the SZ effect in the Bullet-like cluster
  A2146. Crosses indicate the positions of subtracted radio
  sources. The ellipse lower-left indicated the synthesised beam.}
\label{a2146}
\end{center}
\end{figure}

AMI has been used for a variety of different pointed SZ programmes: to
study the astrophysics of individual clusters of particular interest
e.g. the bullet-like cluster Abell 2146 (see Figure~\ref{a2146}) and
the Corona Borealis supercluster; to produce joint constraints on
cluster models once combined with optical lensing data \citep{tash12};
for statistical analyses of scaling relations through observations
towards well-defined samples of clusters drawn from X-ray surveys
(e.g. the LOCUSS, BCS and MACS samples).  We now detail one particular
example as follows.

\begin{figure}[t]
\begin{center}
\includegraphics[width=7.5cm,height=6.5cm,clip=,angle=0.]{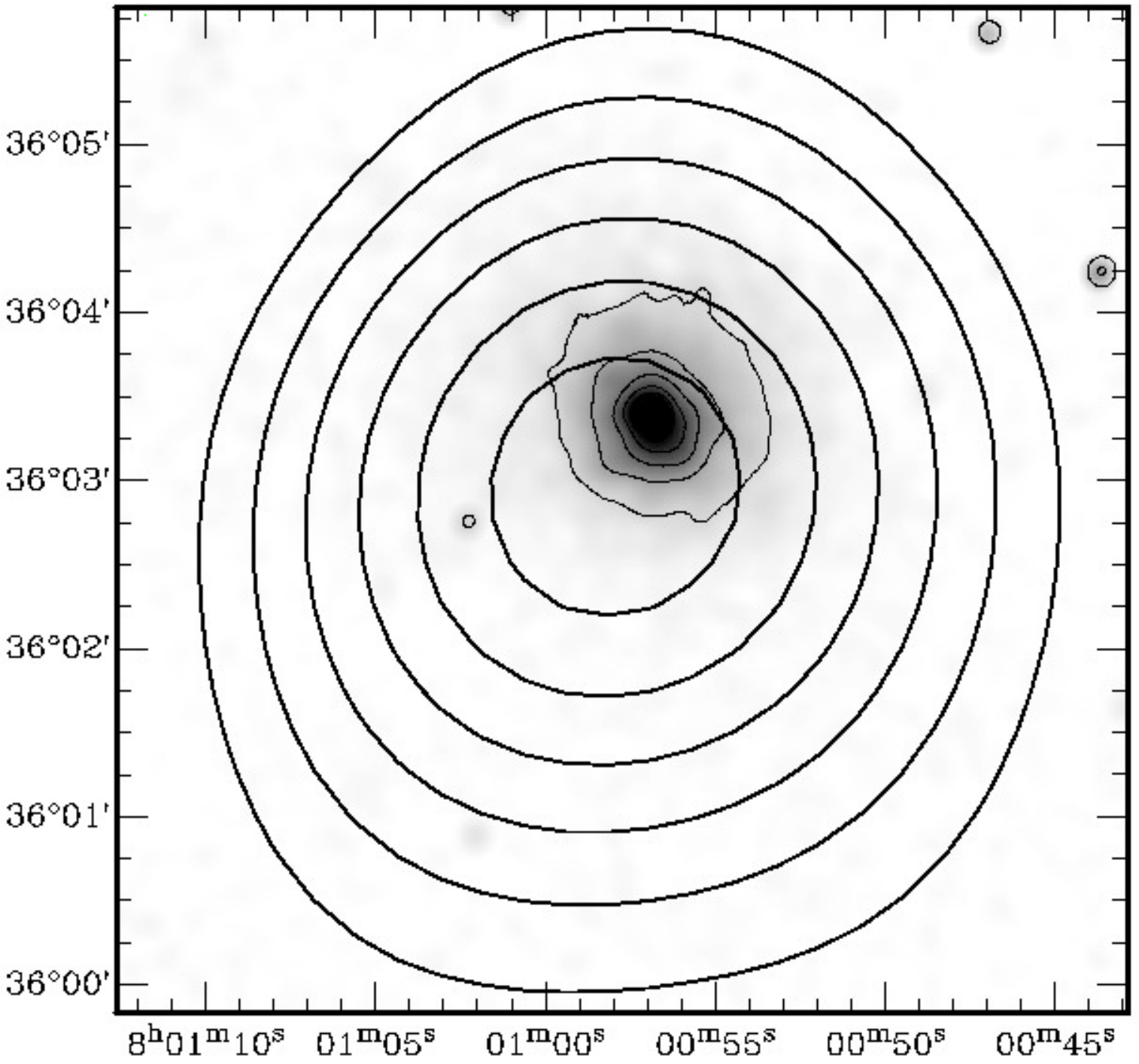}\qquad\includegraphics[width=7.5cm,height=6.5cm,clip=,angle=0.]{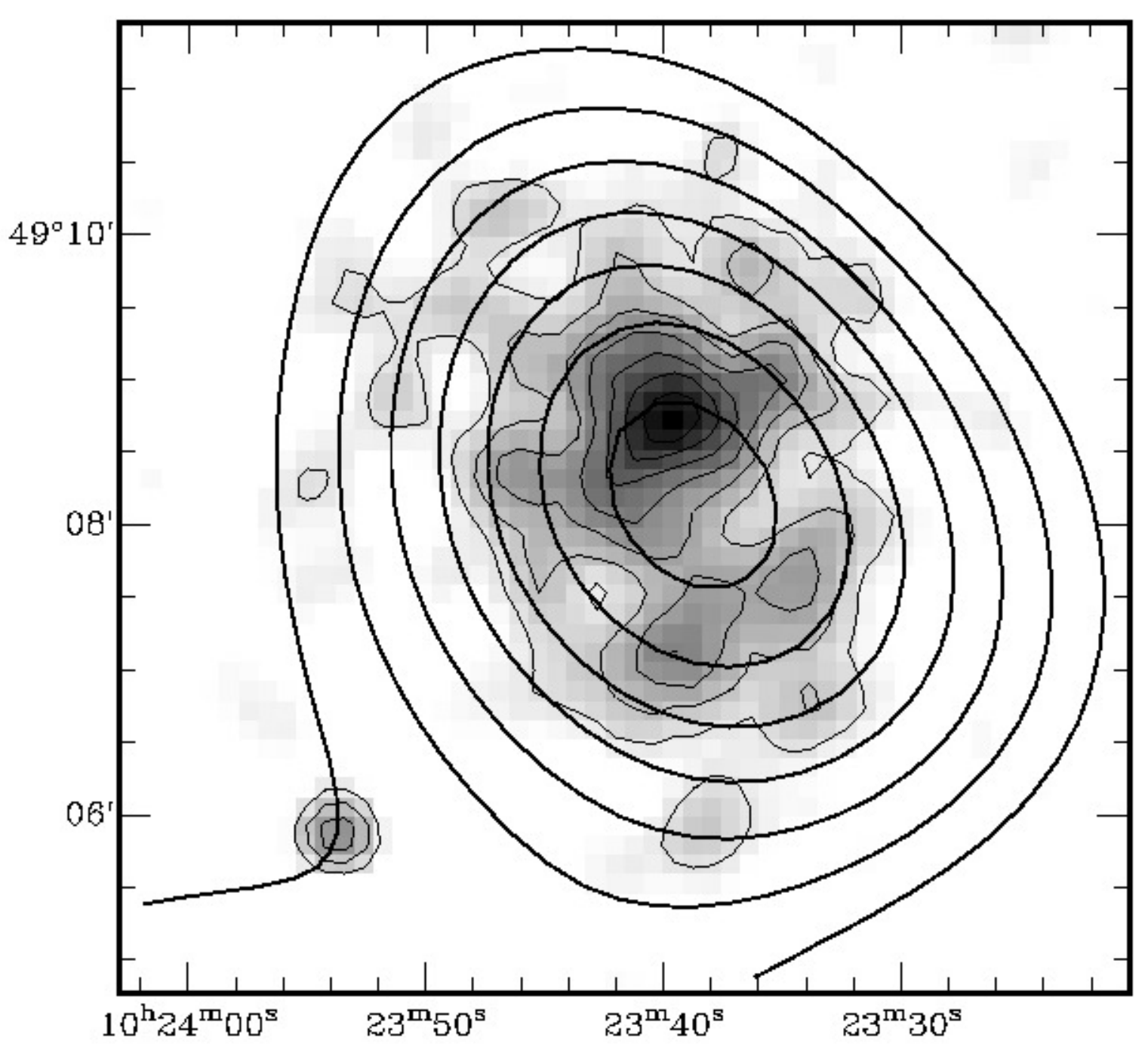}\\
\includegraphics[width=7.5cm,height=6.5cm,clip=,angle=0.]{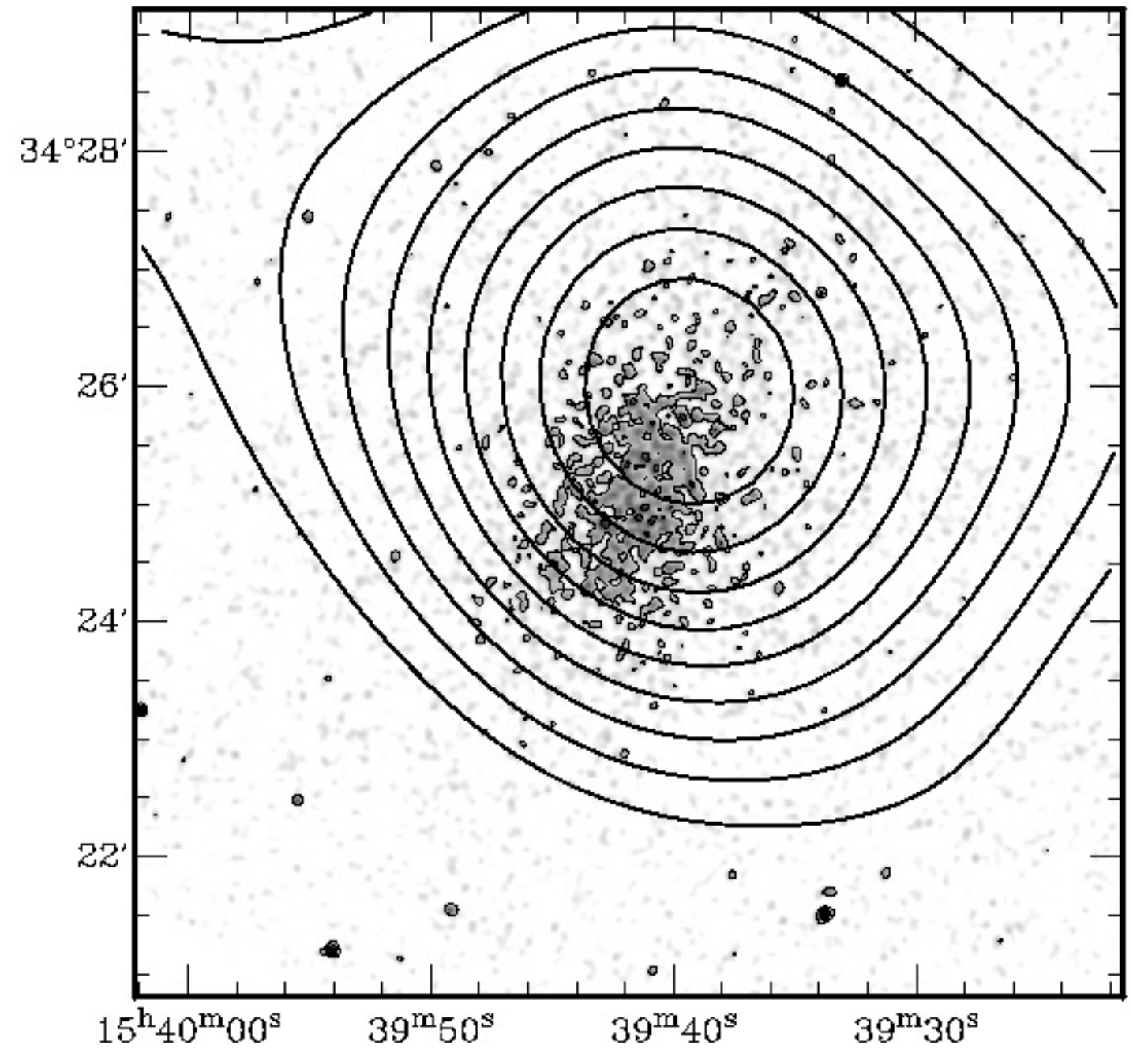}\qquad
\includegraphics[width=7.5cm,height=6.5cm,clip=,angle=0.]{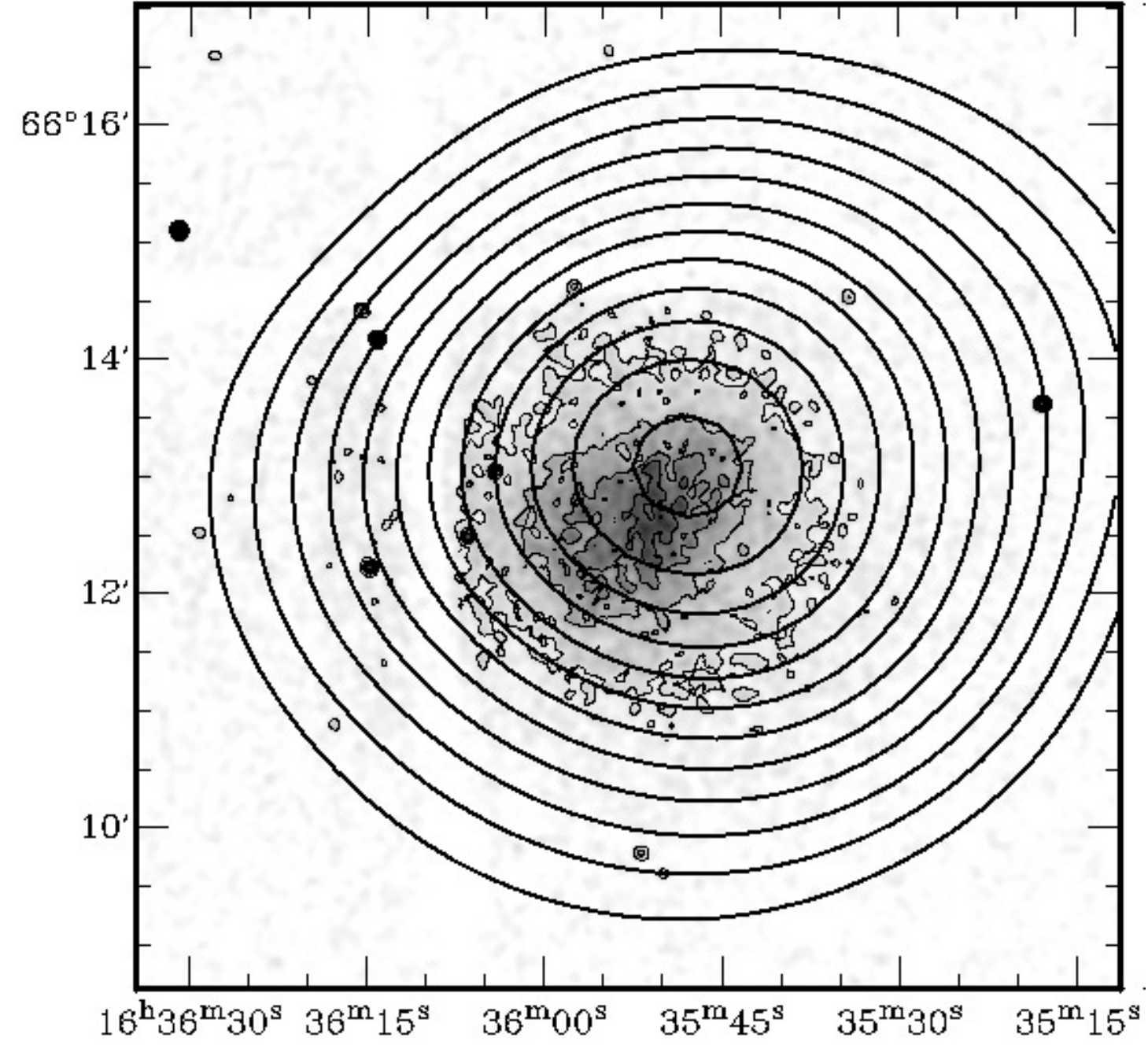}
\caption{Maps of four clusters from the LoCuSS sample; from top left:
  A611, A990, A2111, A2218. Contours show the SZ decrement measured by
  the AMI-SA after source subtraction, while the greyscale is the
  X-ray image smoothed {\it Chandra} data except for A990 which is
  {\sc \textit{ROSAT} HRI}).\label{locuss}}
\end{center}
\end{figure}

AMI observations of an X-ray luminosity-limited subset of 19 LoCuSS
clusters (which have $0.142 \le z \le 0.295$ to minimise cosmic
evolution) have yielded 16 SZ detections with significances (of the
peaks alone) between $5 \sigma$ and $23 \sigma$ (see
Figure~\ref{locuss} for SZ maps overlaid on X-ray for four of these
clusters). The SZ images span scales out to the cluster virial radii
(which very few X-ray images do), and show a larger range of
morphology than that seen in published ACT, SPT or CARMA/SZA images
(which tend not to reach $r_{\mathrm{virial}}$) or in Planck images
(which do not have enough angular resolution).  Under reasonable
assumptions, the cluster plasma temperature can be derived from AMI
data alone \citet{carmen11b} --- unlike the traditional method, no
X-ray information is required. In the 10 cases where deep
\textit{Chandra} or Suzaku images are available, allowing X-ray
temperatures to be measured out to around 0.5$r_{\mathrm{virial}}$
(rather than the more common 0.1--0.2$r_{\mathrm{virial}}$), it is
clear that there is decent correspondence between these $T_{X,
0.5\mathrm{virial}}$-values and the $T_{AMI, 0.5-1\mathrm{virial}}$
values from SZ, but with some strong outliers that have high
$T_{\mathrm{X}}$ compared with $T_{\mathrm{AMI}}$. These outliers are
the strong mergers. AMI easily probes the outer parts of clusters
where most of the baryons and dark matter exist; these parts are hard
for X-ray instruments to detect even at moderate $z$. At moderate $z$,
the combination of AMI and \textit{Chandra}/Suzaku is thus very
powerful for understanding how clusters work.

There are several ongoing collaborations that will allow joint
analysis of our AMI images with data from other wavebands, as
summarised below.

\begin{itemize}

\item {\bf MACS sample} This is a high-redshift sample and so offers
  the possibility of investigating potential evolution in cluster
  properties. Deep \textit{Chandra} data and X-ray temperatures can be
  used to constrain the priors for our {\sc McAdam} analysis.

\item{\bf The Canadian Cluster Comparison Project~(CCCP) sample}. The
  CCCP team have CFHT lensing observations towards a large sample of
  X-ray selected clusters; many have been observed with
  \textit{Chandra} and \textit{XMM}.

\item {\bf The CLASH sample}. The CLASH team has lensing data from
  \textit{HST} and Subaru of a sample of dynamically-relaxed, massive
  clusters and this will allow us to derive stringent constraints on
  dark matter and hot baryons in the representative clusters.

\item {\bf Serendipitous \textit{XMM} cluster detections} \textit{XMM}
  images and temperatures will be available for combination with the
  AMI data.

\end{itemize}

There continue to be huge opportunities for new science from pointed
AMI SZ observations as new cluster samples become available and for
joint analyses with other data sets to explore the physics of the
intracluster gas of particular systems. In addition, there are
possibilities for entirely new AMI SZ science areas, such as attempts
to detect the SZ effect from the plasma in filaments of the cosmic
web, and combination of a AMI-derived gas model with rotation-measure
observations from the new generation of radio observatories such as
JVLA and MeerKAT to measure cluster magnetic fields. One of the most
exciting opportunities in the short term is support of the
\textit{Planck} satellite, which is now providing the first all-sky SZ
cluster survey. This is discussed in the section~\ref{ami-planck}.

\subsection{Comparison of AMI to other SZ telescopes}

The South Pole Telescope (SPT) and the Atacama Cosmology Telescope
(ACT) are multipixel bolometric telescopes specifically designed for
SZ work. They are primarily survey instruments and have already
produced extremely exciting results (see
e.g. \citet{2012arXiv1203.5775R} for SPT and \citet{2011ApJ...737...61M}
for ACT). Both SPT and ACT are conducting very wide field, relatively
shallow surveys and so are detecting the rare, most massive clusters
in the Universe, while AMI's survey strategy is to spend longer
integration times on a smaller area and so sample typical clusters.  AMI
is therefore able to probe the cluster mass function significantly
deeper and so the three telescopes produce useful complementary
information on structure formation. AMI is an excellent instrument for pointed
follow-up observations on samples of galaxy clusters from other
instruments since its 20~arcmin field-of-view matches the typical size
of $z>0.15$ clusters very well; SPT and ACT's very wide fields-of-view mean
that their strength is in survey observations. Finally, both ACT and
SPT are located in the southern hemisphere with the result that they
have little or no (respectively) coverage of the northern sky visible
to AMI.

The SZA is now part of the CARMA telescope and so is no longer a
dedicated SZ instrument. It is similar in many ways to AMI, but AMI
has two key differences which are important for SZ science: 
\begin{itemize}
\item AMI has access to a factor of two times shorter baselines (in
  terms of wavelength, i.e. in the uv-plane), and so samples structure
  out to 10 arcmin, allowing mapping of the gas out to the virial
  radius in galaxy clusters between z=0.15-2.
\item The AMI-LA has ten times the collecting area (and so ten times
  the flux sensitivity) of the AMI-SA and a good coverage of the
  uv-plane. This gives very high sensitivity, high image fidelity
  mapping of the radio point sources, which are the key contaminant
  and dominant systematic for SZ work. For comparison, SZA configures
  two of its eight antennas as outriggers for source subtraction,
  giving comparatively poor flux sensitivity and a single narrow strip
  of coverage in the uv-plane, resulting in a poor point-spread
  function and so poor image fidelity.
\end{itemize}

\section{AMI's role in \textit{Planck} Validation and
Follow-up}\label{ami-planck} 

\subsection{Introduction}

AMI is very useful for \textit{Planck}-related work in two main
contexts --- validation and follow-up.  Due to restrictions on
reporting of \textit{Planck} results before publication, full details
of this work will become available early in 2013.  However, an MoU
between \textit{Planck} and AMI exists covering both these areas, and
a broad description of the type of work involved will now be given.

\subsection{Scientific Context}

\begin{figure}[h]
\includegraphics[width=6cm]{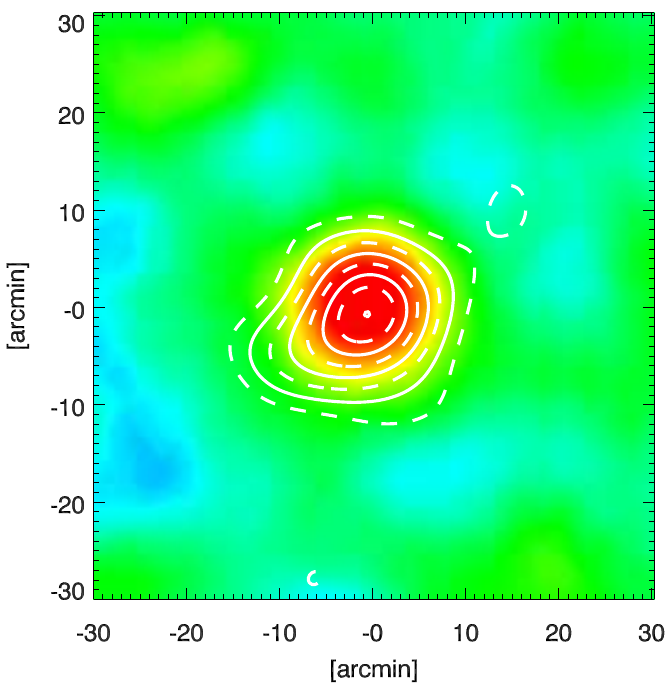}
\caption{y-map of PLCKESZ G139.59+24.19 as observed by \textit{Planck}
  (colour image) and AMI (contours) at a common resolution of 13
  arcmin. The contours are from two to nine in S/N ratio. (Taken from
  \citet{2011A&A...536A...8P}.)\label{fig:ami-planck}}
\end{figure}

Last year saw the first release of \textit{Planck} results for the SZ
effect in clusters of galaxies. The resolution of \textit{Planck} for
SZ studies (at best $\simeq 5'$) is lower than for most ground-based
observations, but is compensated for by all-sky coverage, plus a good
frequency discrimination of the effect due to multiple frequency
channels spanning the whole range over which it is important. The
`Early Release SZ Catalogue' (ESZ) \citep{2011A&A...536A..10P}
contained a sample of 189 clusters of which 169 were previously
known. Ground-based telescopes have now confirmed the SZ detections
for some of the cluster candidates; AMI provided confirmation within
the ESZ paper itself for a previously unknown cluster (see
Fig.~\ref{fig:ami-planck}), and has subsequently done the same for a
further cluster \citep{tash11}, as well as providing refined position
estimates.

There is great cosmological interest in the factor of between 0.5 and
0.7 by which the SZ decrements measured by \textit{WMAP7} were lower
than what was expected for the given clusters using standard X-ray
models for temperature and profile \citep{2011ApJS..192...18K}. Some
support for a deficit by this same factor came from South Pole
Telescope measurements of the SZ power spectrum at high $\ell$
\citep{2010ApJ...719.1045L}. However, one of the first \textit{Planck}
analyses \citep{2011A&A...536A..10P} has found no evidence for a
deficit in SZ signal strength in \textit{Planck} data relative to
expectations from the X-ray properties of clusters, in apparent
contradiction with the \textit{WMAP} results, and highlighting the
need to be able to incorporate independent high-resolution SZ data. In
\citet{2011A&A...536A..10P}, in addition to emphasising that no
deficit has been found, it is pointed out that, even when the analysis
is redone in two bins with pressure profiles corresponding to `cool
cores' or `morphologically disturbed' types, the results are still
robust to this, with maximum deviations at only the few percent
level. Since this division was thought to be important in explanation
of the \textit{WMAP7} results \citep{2011ApJS..192...18K}, it is
unclear currently why there is the apparent discrepancy in findings,
though a linkage with assumed profile shape still seems likely. More
generally, the comparison of SZ radial profiles with X-ray profiles is
an important part of the study of clusters, particularly for the outer
regions where AMI excels.

\subsection{AMI Role in Validation and Follow-up}

An advantage of SZ verification of candidate clusters discovered in an
all-sky satellite survey, is that SZ surface brightness is independent
of redshift. This means that even for a high-redshift cluster in which
IR, optical or X-ray verification may fail, direct SZ verification is
still possible. Detections of this kind, of a candidate high-redshift
cluster, are potentially some of the most exciting individual objects
that \textit{Planck} will detect, and thus need a high level of
reliability associated with them. AMI has the sensitivity to be able
to confirm such a detection on a timescale that can make it useful
during the verification cycles. Additionally, by performing checks of
the SZ levels over a statistically selected sub-sample of all
\textit{Planck} SZ detections, a check can be made of the photometric
accuracy of the \textit{Planck} SZ amplitude. AMI is also able to
provide smaller positional uncertainties on candidates than
\textit{Planck} alone, and this feature is also useful in the process
of cluster follow-up and validation with optical and X-ray telescopes.

Validation observations can be interleaved with a programme of deeper
observations ($\simeq 32$\,hr/cluster) on selected clusters. For rich
clusters this will give well resolved $\simeq 20\,\sigma$ detections,
including the outer regions of the cluster gas, and provide radio
spectra in the 13.5--18.0\,GHz range in six spectral channels (which
may be useful for constraining the level of possible cluster
synchrotron halo emission in combination with data from LOFAR and
KAT7). AMI-LA point-source subtraction is able to keep up with the
demands of the AMI-SA SZ observations in real time, so these integration
times include the time needed to identify and remove contaminating
point sources.

The AMI/\textit{Planck} teams have worked on a pilot study involving
detailed comparison between SZ measurements obtained by both
\textit{Planck} and AMI for a first sample comprising 11 clusters
\citep{arXiv:1204.1318}. This involved detailed numerical simulation
of the respective responses of the two instruments to a common cluster
profile. Since \textit{Planck} and AMI are sensitive to different
scales, their comparison and/or combination is already revealing
interesting new cluster physics. For example, is the ``Universal
pressure profile'', commonly adopted to describe galaxy clusters
\citep{2010A&A...517A..92A}, sufficient to explain simultaneously the
\textit{Planck} and AMI measurements or is there more complicated
physics at play? Additionally, the different sensitivities of Planck
and AMI to different regions of scale space, mean it is possible to
investigate the potential to break parameter degeneracies, e.g. in the
$Y_{SZ}-\theta$ plane, even in an SZ-only analysis. The results of the
analysis show an overall broad agreement between the AMI and Planck
results but with a tendency for the AMI fluxes to be smaller,
suggestive indeed of a possible mismatch in the assumptions about
cluster profile.

AMI has continued to play a significant role in Planck cluster
validation for a much larger sample of potential new clusters, and
extension of the comparison between the Planck and AMI results
to the entire sample of clusters visible to AMI in the forthcoming
Planck catalogue is currently underway. Ultimately, constraints from
other cluster probes (e.g.\ weak lensing and X-ray measurements) can
then also be investigated --- in principle the independent
constraints within the parameter likelihood planes taking different
pairs of methods, should all be consistent, thus providing a stringent
test of whether the cluster model and observations are all correct.

\section{Galactic research with AMI}\label{galactic}

\subsection{Introduction}
\label{sec:intro}

The microwave band is a highly under-utilised spectral window for
Galactic research. Historically this is because the expected emission
has relatively low surface brightness compared to the
longer-wavelength non-thermal Galactic emission and the
shorter-wavelength sub-mm and infra-red blackbody emission from
thermal dust; emission in the microwave band is predominantly due to
the thermal bremsstrahlung, or free-free, mechanism. However, given
sufficient sensitivity the microwave band offers a \emph{unique}
window on numerous vital physical processes including, but not limited
to, the formation of stars in both the high and low mass regimes, as
well as the earliest stages of planet formation from circumstellar
disks. These processes are particularly well probed at
centimetre-wavelengths as it is here that opacity effects in the hot
ionized plasma around young stars are most informative and here that
opacity effects in dust populations are most reduced. This historical
observational bias against microwave wavelengths may also be the cause
of the late discovery of the anomalous microwave emission largely
attributed to spinning dust.

The AMI telescope provides a highly effective instrument for Galactic
science at centimetre wavelengths, not only for its high sensitivity
in the microwave band, but also for its extensive spatial dynamic
range. Galactic emission exists on a broad range of scales with
physical processes often strongly linked across that range. With high
sensitivity and resolution from 0.5-10\,arcmin AMI is ideal for
probing Galactic processes on multiple scales \emph{simultaneously}.
A range of successful galactic science programmes have already been
carried out by AMI.  Here we review the major components of this
science programme with an emphasis on where the future direction of
the AMI telescope may take us in these areas.

\subsection{Spinning Dust}

\subsubsection{Scientific Motivation}

Anomalous Microwave Emission (AME) was termed by
\citet{1997ApJ...486L..23L} who detected dust-correlated emission at
15\,GHz that could not be explained by traditional emission mechanisms
(e.g. CMB, free-free, synchrotron, thermal dust). It was also easily
detected by the COBE-DMR satellite \citep{1996ApJ...464L...5K} but was
initially thought to be due to free-free emission from warm ionized
gas. Since then, numerous experiments have measured AME across the sky
over the frequency range 8--100~GHz (see \citet{2011A&A...536A..20P}
and references therein). There has been much debate about the exact
nature of the AME. A number of mechanisms have been proposed including
hot free-free emission, hard synchrotron radiation, two-level system
dust and magneto-dipole radiation (see e.g.
\citet{2003ApJS..148...97B} for a discussion). However, the leading
theory for AME is electric dipole radiation from small rapidly
spinning dust grains. It has been known for a long time
\citep{1957ApJ...126..480E} that residual electric dipole moments of
small dust grains could potentially emit detectable radio
emission. However, it took 50 years for it to be detected because it
appears to emit strongly in a relatively narrow range of frequencies
(10-60\,GHz). The theory is now relatively well understood
\citep{1998ApJ...494L..19D,1998ApJ...508..157D} although the exact
details of excitation and damping of the dust grains are
complicated. Statistically, on large scales ($\sim 1$\,degree) the
latest results from \textit{Planck} now appear to show definitively that
spinning dust emission is indeed responsible, at least for some
lines-of-sight \citep{2011A&A...536A..20P}. The study of AME is now a
significant area of astronomical research. Not only is it a strong
foreground that must be accurately removed when measuring CMB
anisotropies, particularly at frequencies near and just below the
foreground minimum at $\simeq$70\,GHz (\citep{2006cmb..confE..18D}),
and consequently understanding the spectrum of AME and mapping its
morphology across the sky will aid component separation, leading to
more precise measurements of the CMB; but moreover, spinning dust
offers a new window into the physics of dust grains and their environs
with particular emphasis on the small grain population which is
crucial for controlling the physical conditions and chemistry of star
forming regions. Measuring its spectrum and spatial morphology on
smaller scales provides a new way to study dust that is complementary
to traditional methods that typically use infrared continuum and
spectroscopy.

\subsubsection{Contribution of AMI}

AME has been observed across the sky by numerous experiments. However,
the majority of these have been CMB experiments that work at higher
frequencies ($> 20$\,GHz) and at relatively low angular resolution
(5~arcmin). On large angular scales the most important dataset has
been from the WMAP satellite, due to its lowest two frequency channels
at 23\,GHz and 33\,GHz. New data from the \textit{Planck} satellite is
already producing astonishing new results but the data are limited to
its lowest frequency channel at 30\,GHz with 30~arcmin
resolution. With satellites such as WMAP and \textit{Planck} limited
to large angular scales and high frequencies, it is therefore crucial
to survey the sky at higher angular resolution and at frequencies
below those accessible to \textit{WMAP} and \textit{Planck} to gain a
complete understanding of this emission. The AMI instrument is
therefore perfectly matched for studying AME for multiple reasons: its
frequency range (12-18\,GHz) covers the low side and peak of spinning
dust emission; its high angular resolution (2~arcmin for AMI-SA)
allows it to be well-matched to radio surveys at lower frequencies and
to localise emission, whilst the size of its primary beam (20~arcmin)
is well matched to CMB experiments at higher frequencies, including
\textit{Planck}; the compact nature of the array results in high surface
brightness sensitivity and consequently high spatial dynamic range
mapping; when combined with the Large Array it is able to investigate
and separate the source population with high precision, which is
especially valuable in star forming regions.

\begin{figure}
\centerline{\includegraphics[width=0.4\textwidth]{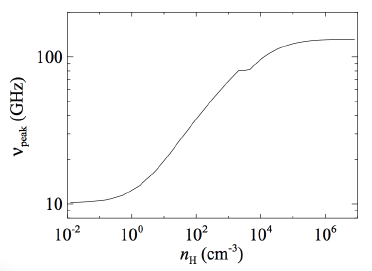}}
\caption{Variation of the peak of the spinning dust spectrum as a function of
  density \citep{2009MNRAS.395.1055A}. Lower density
  environments peak at $<20$\,GHz and are accessible to
  AMI.\label{fig:dust-spec}} 
\end{figure}

These benefits have been demonstrated to good effect by the
significant contribution to spinning dust research that AMI has
already made
\citep{2008MNRAS.385..809S,2009MNRAS.394L..46A,2009MNRAS.400.1394A,2010MNRAS.403L..46S,2010MNRAS.406L..45S}. In
addition a recent survey of the Perseus molecular cloud region is
showing spectacular results. The detailed structure and comparison
with infrared tells us that the spinning dust critically depends on
location and environment \citep{2009MNRAS.395.1055A}.  More
observations of diffuse clouds are needed to unravel this complicated
picture and relate it to theoretical predictions and AMI is ideally
suited to these kinds of surveys. In addition, \textit{Planck} is
already detecting new AME candidates that will need follow-up at high
resolutions. AMI data can confirm whether the AME is indeed diffuse
emission or if the large \textit{Planck} beam is contaminated by
compact HII regions that are emitting optically thick emission at high
($>10$\,GHz) frequencies. The frequency coverage of AMI can pin down
the peak of the spinning dust spectrum, which in the most diffuse
regions appears to be below 20\,GHz. In the more dense regions, AMI
can measure the rise at low frequencies due to the larger grains and
the data are complementary to that at higher
frequencies. Fig.~\ref{fig:dust-spec} shows how the peak of the
spinning dust curve, which can be as low as 10~GHz, varies with
density. Although other arrays can probe this frequency range
(e.g. JVLA, ATCA), they do not have the brightness sensitivity required
to measure diffuse emission on arcminute scales.

\subsection{Low Mass Star Formation}

\subsubsection{Scientific Motivation}

Since the advent of the \textit{Spitzer} space telescope, direct
infrared detection of the hot dust heated by an embedded protostar has
become the most popular method for differentiating between starless
and protostellar cores. Although the method is extremely reliable,
\textit{Spitzer} identification is not completely certain
\citep{2009A&A...502..139H} as the response to extragalactic radio
sources is known to mimic that of embedded protostellar cores and can
cause false detections. In addition, a protostellar spectrum alone
will not demonstrate that a core is truly embedded and, like many
surveys, \textit{Spitzer} is also flux-limited, with a luminosity
completeness limit of $0.004(d/{140~{\rm pc}})^2$ ${L}_{\odot}$
\citep{2008ApJS..179..249D,2007ApJ...663.1149H}. Correctly determining
the relative numbers of starless and protostellar cores in
star-forming regions is essential for inferring timescales for the
different stages of protostellar evolution; for low-luminosity objects
it is also necessary to determine the extent of the luminosity problem
first articulated by \citet{1990ApJ...349..197K} in order to
understand the accretion mechanisms underlying protostellar evolution.

Identifying embedded protostars by detecting their molecular outflows,
either via high-excitation jet interactions (Herbig-Haro objects and
shock-excited H$_2$;
e.g. \citet{2008MNRAS.387..954D,2005AJ....129.2308W}) or through
low-excitation molecular lines such as $^{12}$CO, can immediately
identify a source as an embedded protostar, avoiding the contaminants
of infrared colour selection. However, in crowded star-formation
regions, confusion from neighbouring outflows can frequently be an
issue.  Radio emission (see e.g. \citet{1993ApJ...406..122A}) is a
reliable method for detecting protostars, as well as providing a
potential mechanism for distinguishing protostellar class via its
correlations with other physical characteristics.

\begin{figure}
\centerline{\includegraphics[width=0.5\textwidth]{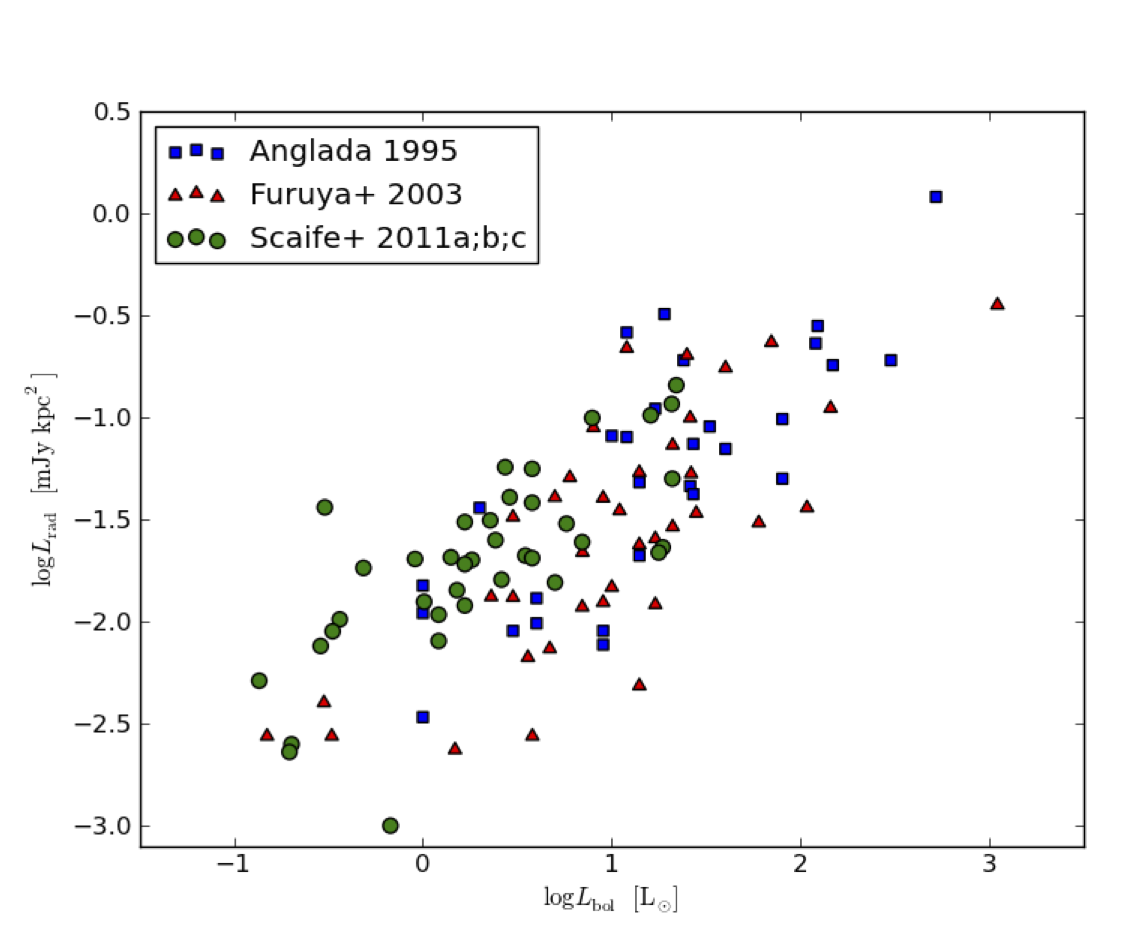}}
\caption{\label{fig:lbol} Correlation of radio luminosity with bolometric
  luminosity for Class~0 \&~I YSOs} 
\end{figure}

\subsubsection{Contribution of AMI}

Radio follow-up of the \citet{2008ApJS..179..249D} \textit{Spitzer}
catalogue of low luminosity embedded objects at 15.7\,GHz
\citep{2011MNRAS.410.2662A,2011MNRAS.415..893A,2012MNRAS.420.1019A}
(see Fig.~\ref{fig:lbol}) has provided clear evidence for distinct
trends in the behaviour of the radio luminosity of these sources when
compared with their bolometric luminosity, IR luminosity and outflow
force where molecular outflows have been measured. Similar trends are
well-documented at lower radio frequencies for higher luminosity
sources ($<10^3$\,L$_{\odot}$; see e.g. \citet{1995ApJ...443..682A}
or \citet{2007ApJ...667..329S}).  In addition, the relationship between
radio luminosity and the IR luminosity of these objects has shown that
it may be possible to use the radio emission as a proxy for the
internal luminosity of low luminosity objects, a quantity which can
otherwise only be derived through complex modelling of the IR spectra.
The cm emission probes the current mass ejection directly, without
the historical averaging inherent in CO outflows, and originates from
the inner protostellar core which will be the subject of ALMA and
EVLA studies.  The cm datapoints on the SED provide constraints both
on the free-free emission geometry and on the relative contributions
of thermal dust and bremsstrahlung \citep{2012MNRAS.423.1089A}.

Blind surveys for radio protostars will be intrinsically biased
towards young objects \citep{2012MNRAS.420.1019A} and the radio
emission from heavily embedded sources may provide an excellent method
for confirming the Class~0 population. This is of particular
importance in light of the newly discovered VeLLO objects, many of
which are still heavily embedded. Such objects are difficult to
confirm in the infra-red \citep{bourke2006} or through molecular
mapping \citep{bourke2005, crapsi2005}, but nevertheless represent an
important population for theories of non-steady accretion. AMI
provides an important resource for studies to confirm and characterize
low luminosity sources. As has already been demonstrated by the
instrument, the cm-wave band is an excellent probe of protostellar
activity. Blind surveys of molecular clouds with the AMI-LA that are
currently in progress will also provide key information on young
embedded objects which may previously have been missed by surveys at
higher frequencies.

\subsection{High Mass Star Formation}

\subsubsection{Scientific Motivation}

In recent years we have started to realise that the early lives of
young HII regions (the ultracompact \& hypercompact phases) may be
dynamic epochs in the star formation process, with complex accretion
flows through the ionisation boundary \citep{2011MNRAS.416.1033G} that
are often mediated by Rayleigh-Taylor instabilities
\citep{2011ApJ...732...20K}. The classical picture of young HII
regions is one of simple pressure-driven expansion whereby the central
star has finished accretion, has reached the main sequence, and the
surrounding HII region is expanding at the sound speed into the
surrounding ISM. However, recent observations of ultracompact and
hypercompact HII regions reveal that often infall and accretion are
ongoing in both the surrounding molecular
(e.g. \citet{2009ApJ...703.1308K,2011A&A...532A..91B}) and ionised gas
\citep{2006Natur.443..427B,2009ApJ...706.1036G} and more
intriguingly, some ultracompact and hypercompact regions show evidence
of variability on timescales of a year
\citep{1998ApJ...495L.107A,2004ApJ...604L.105F,2009ApJ...706.1036G}.

Detailed modelling of HII region variability has been carried out by
\citet{2010ApJ...719..831P,2011MNRAS.416.1033G}, who find that the
radio flux of young HII regions changes in direct response to sudden
increases in the accretion of material onto the star. As the accretion
rate increases the HII region reduces in size and is quenched. Thus
tracking the time-variable radio flux from the ultra and hypercompact
HII regions in the Peters et al. model enables the accretion rate onto
the star to be inferred, and offers the unique prospect of determining
the accretion history. By comparison, to determine the accretion
history via spectroscopic measurements of the small-scale velocity
structure (a few hundred AU) around the star would require angular
resolution better than $\sim$10 milli-arcsec --- a challenging target
for interferometers. Determining the accretion history for massive
star formation will allow current star formation models
(e.g. \citet{2003ApJ...585..850M,2011MNRAS.410.2339B}) to be
confronted directly.

\subsubsection{Contribution of AMI}

\begin{figure}
\centerline{\includegraphics[angle=-90, width=0.4\textwidth]{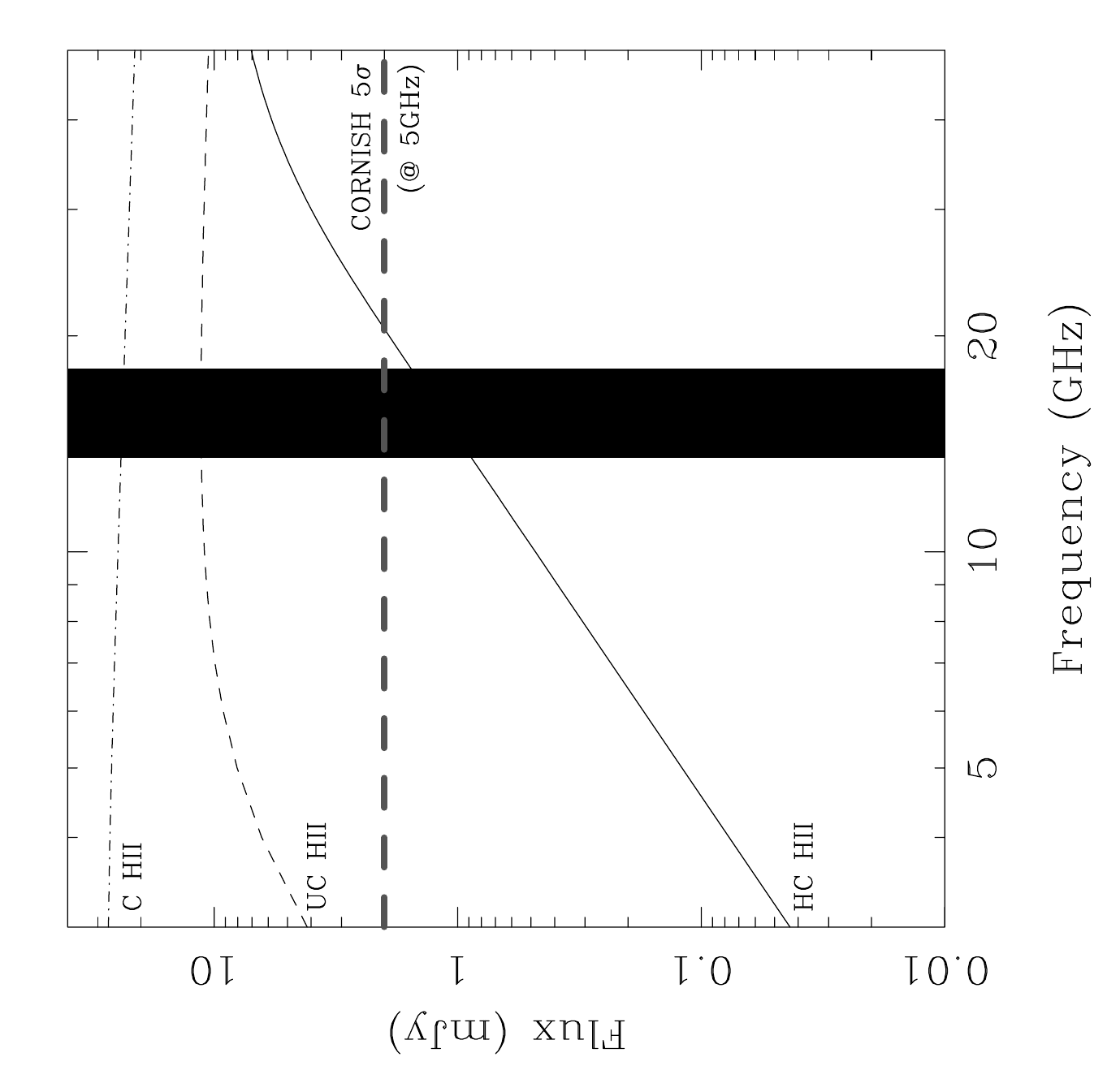}}
\caption{Theoretical spectra of isothermal young homogeneous HII regions at a
  distance of 20 kpc. Solid, dashed and dot-dashed lines show spectra for
  hypercompact, ultracompact and compact HII regions respectively, with
  typical emission measures taken from \citet{2005IAUS..227..111K}. We show
  the 5GHz 5$\sigma$ flux limit for the CORNISH survey by a horizontal line,
  and the frequency range of AMI with a solid vertical bar. Operating at 15.7
  GHz will enable us to study ultracompact, hypercompact and compact HII
  regions.}
\label{fig:hiispec}
\end{figure}

AMI is of fundamental value in this kind of study. Firstly, the
operating frequency of 15.7\,GHz is close to the peak in the spectrum
of most ultracompact HII regions and takes full advantage of the steep
positive spectral index of hypercompact HII regions
(e.g. \citet{2005IAUS..227..111K}). This is shown in
Figure~\ref{fig:hiispec}, which shows the SED of ultracompact,
hypercompact and classical HII regions. Secondly, AMI is an excellent
telescope for carrying out variability studies as the system and
calibration are both stable and well-understood. Finally, AMI has
sufficient flexibility in its schedule to enable variability studies
with a range of cadences. It will be important to identify isolated
ultracompact and hypercompact regions for variability studies with
AMI, due to the 30$^{\prime\prime}$ beam of the Large Array and the
typical clustering of massive star forming regions. Such isolated
sources can be identified from the northern section of the CORNISH
survey (for ultracompact HII regions only,
\citet{2010HiA....15..781P}), the Torun and MMB methanol maser surveys
(by searching for associated hypercompact HII regions,
e.g. \citet{2007MNRAS.379..535L}), and the \emph{Herschel} Hi-GAL
survey \citep{2010A&A...518L.100M}.  Conducting an AMI variability
study on isolated young HII regions in the North will not only give us
a greater understanding of the accretion processes involved in massive
star formation, but will also feed into the much larger statistical
studies possible at sub-arcsecond resolution with MeerGAL, the MeerKAT
14\,GHz Galactic Plane Survey.

The high sensitivity of the AMI-LA has already been demonstrated as
providing an excellent tool for long term monitoring programmes of
{\sc hchii} regions (AMI Consortium: Scaife et~al., in prep.) using
total flux density as a tracer of expansion and contraction for
optically thick emission. In addition the AMI Galactic Plane Survey
will provide a unique resource in the Northern hemisphere for
detecting new {\sc hchii} regions due to its large area coverage (860
square degrees) and depth (AMI Consortium: Perrott et~al., in prep.).

\subsection{Planet Formation}

\subsubsection{Scientific Motivation}

The growth of dust grains from sub-micron to micron sizes in the local
environment of protostellar objects is well-probed by observational
data in the mid-infrared; however, the further growth of these grains
to larger sizes is only probed by their sub-mm and cm-wave
emission. It is this growth, and the possible coagulation and
sedimentation that follows it, which is widely thought to lead to the
formation of proto-planetary objects in the dense disks around
evolving protostars. Measuring the long wavelength tail of the thermal
emission from dust, which is expected to be dominated by such large
dust grains, often referred to as pebbles, provides an independent
measure of the dust composition in the disk through its opacity index
($\beta$), as well as the mass of the accumulated matter in the disk
which is an important factor in models of planet formation
(e.g. \citet{boss1998}).

Sub-mm measurements towards circumstellar disks are often used to
determine the mass of proto-planetary disks and hence their potential
for planet formation. Alternative methods of disk mass estimation,
such as spectral line measurements of molecular gas, are complicated
by opacity effects (e.g. \citet{1993ApJ...402..280B}) and require
complex models to account for these effects as well as those of
depletion \citep{2003A&A...402.1003D,2004ApJ...615..991K}. Sub-mm and
radio measurements are useful not only as a probe of the disk mass
itself, but, with multi-wavelength data available, they can also be
used to determine the evolution of the opacity index as a function of
frequency \citep{2011ApJ...728..143S,2011AJ....141...39S} and place
constraints on the growth of dust grains in such disks. This in turn
will show whether the timescales assumed in current models of planet
formation are consistent with observational data. However, comparison
of disk evolution models and sub-mm data indicate that in $10^5$\,yrs
dust grains will have already grown to a size large enough to cause
substantial opacity effects and hence problems in estimating disk
masses on the basis of sub-mm data alone. Consequently, it is likely that
sub-mm data can only provide reliable disk masses for Class~0
objects. These values are subject to caveats in the context of further
planet formation as at this evolutionary stage there is also expected
to be significant accretion from the disk onto the central object
\citep{2011MNRAS.412L..88G}. Indeed, to date the disk masses
determined from sub-mm data appear to be too low to agree with
theoretical models of planet formation (see e.g.
\citet{2005ApJ...631.1134A}).

\subsubsection{Contribution of AMI}

\begin{figure}
\centerline{\includegraphics[width=0.6\textwidth]{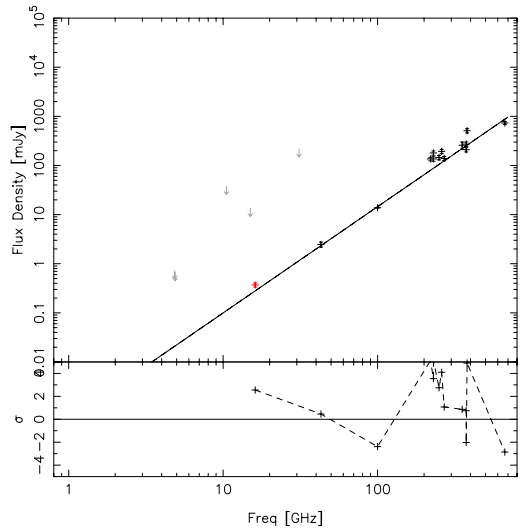}}
\caption{SED of DO\,Tau (Class~II). Black data points are
  taken from the literature, red data point from the AMI-LA, and grey arrows
  indicate upper limits from the literature (AMI Consortium: Scaife et~al., in
  prep.). \label{fig:dotau} } 
\end{figure}

At the resolution of the AMI-LA ($\simeq 25$\,arcsec) it is not
possible to resolve the structure of proto-planetary disks. However,
at cm-wave frequencies the contribution from the larger dust envelope
around proto-planetary systems has a negligible contribution to the
flux density (typically $<1\,\mu$Jy) for the Class~I objects where
planet formation is thought to initiate; consequently the need for
high resolution to separate the disk and envelope contributions, which
arises at sub-mm wavelengths, is not as pressing. At the observing
wavelength of AMI $\lambda \simeq 2$\,cm it is also the case that the
disk emission will be optically thin in the vast majority of cases,
negating the optical depth considerations which affect the disk mass
estimates made at higher frequencies. Consequently using cm-wave data
to estimate disk masses provides a method which is far more applicable
for the more evolved Class~I~\&~II objects where planet formation is
starting to occur. This has already been demonstrated as a proof of
concept with AMI using a sample of disks in the Taurus molecular cloud
\citep{2011arXiv1111.5184S} where cm-wave disk masses were found to be
systematically higher than those derived from sub-mm data, boosting
them above the minimum-mass threshold required for giant planet
formation. Future work with AMI in this area has the potential to
better constrain not only theories of planet formation through disk
mass evaluation, but also, through placing limits on the opacity index
of the disk material, the evolution of the grain growth within such
disks, see Fig~\ref{fig:dotau}. These data are very different from and
highly complementary to those from the JVLA, as well as the ALMA
telescope, particularly the proposed Band 1 instrument at 30\,GHz.

\section{Transient and Variable-Source Astrophysics}\label{transients}

\subsection{Introduction}

AMI-LA, mainly in its former guise as the Ryle Telescope~(RT), has
been an extremely successful telescope in providing key radio
monitoring and follow-up data on a number of high-energy astrophysical
sources, e.g.~X-ray binaries~(XRBs) and gamma-ray bursts~(GRBs). This
has been an extremely productive area in terms of publications and
citations (for example, there are over 3000 citations for papers by
Fender and/or Pooley from this).

In the following, we provide examples of some of the key results over
this period, and then discuss what AMI can do in this area in the
future.

\subsection{Key AMI-LA / RT Results on Transients}

The AMI-LA has been used extensively to monitor currently active XRBs,
systems that contain an accreting neutron star (NS) or black hole
(BH). These systems are the sites of some of the most extreme
astrophysics in the Universe, and black-hole accretion has been shown
to be mass-scalable such that we can learn about AGN accretion and
feedback from studying BH XRBs (and vice versa).

A handful of systems are `persistently active', including several
famous XRBs such as Cygnus X-1, Cygnus X-3, GRS 1915+105 and SS433.
The AMI-LA datasets on these objects, in particular the first three, are
simply the most comprehensive datasets in existence on the variability
of relativistic jets (which produce the radio emission) and their
connection to the accretion flow (as observed in X-ray and infrared).

In addition, there are a large number of transient systems, notably
XRBs in outburst, and GRB afterglows, which show transient radio
emission for a period of typically a few months.

\subsubsection{Cygnus X-1}

\begin{figure}[t]
\begin{center}
\includegraphics[width=11.0cm]{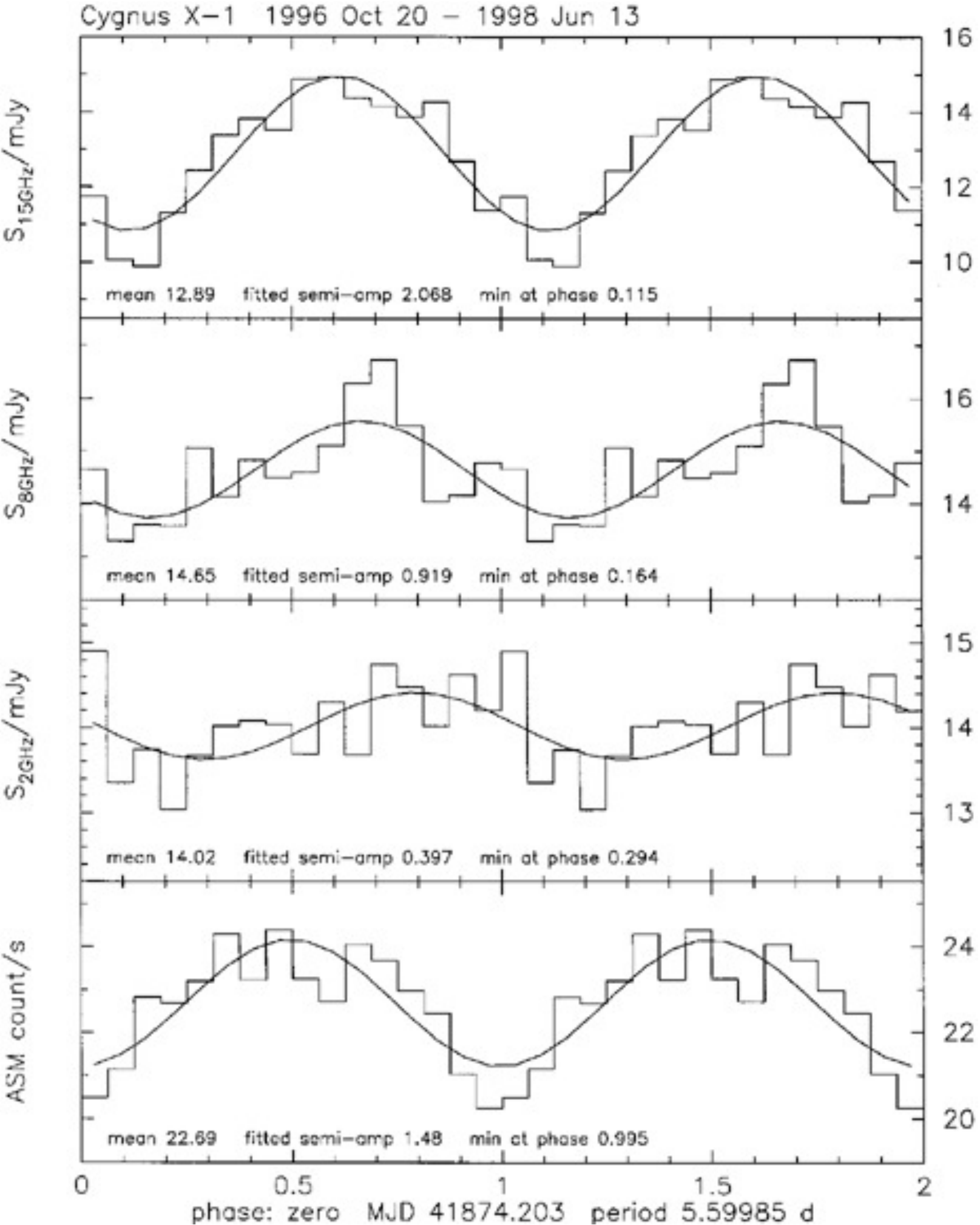} 
\caption{Orbital modulation of the 15.7-GHz radio emission from the
  black-hole binary Cygnus X-1. From \citet{1999MNRAS.302L...1P}.}
\label{cygx1}
\end{center}
\end{figure}

Cygnus X-1 is the archetypal `black-hole candidate', and contains a
$\simeq 10$\,M$_{\odot}$ BH in orbit around an OB-type companion
star. The source accretes persistently around $\simeq 2$\% of its
Eddington luminosity, nevertheless undergoing accretion state changes,
as revealed by occasional strong changes in the X-ray spectral and
timing behaviour (see Figure~\ref{cygx1}).

The AMI-LA monitoring of Cygnus X-1 has provided a key data set for
understanding the jet in this system, which has been demonstrated to
be rather powerful in the long term \citep{2005Natur.436..819G} and
{\em possibly} associated with a the black hole which is possibly
rapidly spinning \citep{2011ApJ...742...85G}.

One of several key discoveries of the AMI-LA for Cygnus X-1 is that
the radio emission is modulated at the 5.6-day orbital period
\citep{1999MNRAS.302L...1P}. This is a unique situation in the whole
of astrophysics: the regular passage of the inner regions of a
relativistic jet moving through the dense parts of the OB star stellar
wind provide variable free-free absorption along the line of
sight. These data provide rich constraints on the location and size of
the relativistic jet (see e.g. \citet{2011MNRAS.416.1324Z}).

\subsubsection{Cygnus X-3}

Just as for Cygnus X-1, the Cygnus X-3 data taken by the AMI-LA and RT
over more than 30 years of monitoring are a unique resource for
understanding jets. The data have generated many tens of papers.

\begin{figure}[t]
\begin{center}
\includegraphics[width=11.0cm]{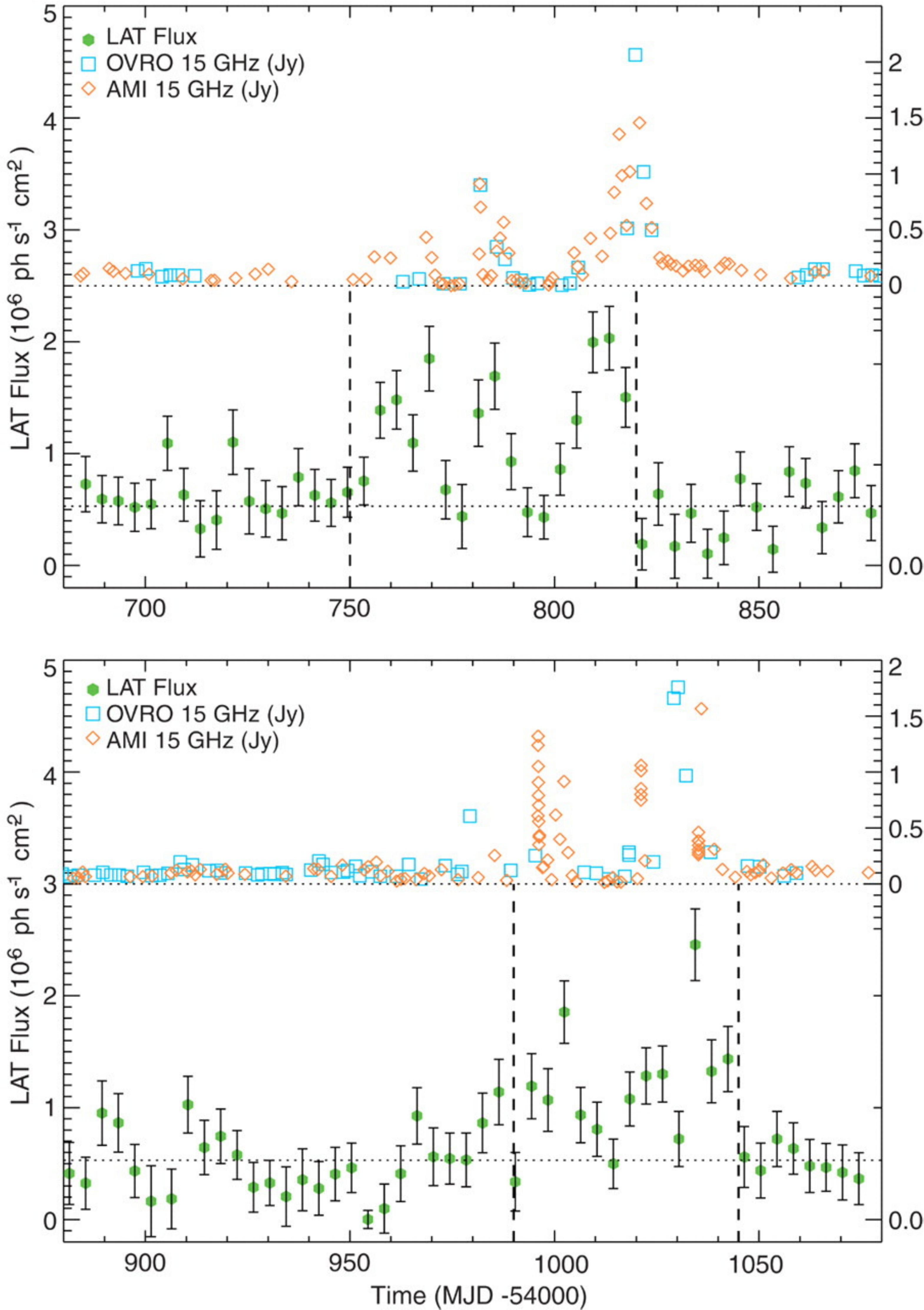} 
\caption{
  Detection of gamma-ray emission coincident with periods of radio
  flaring in the XRB Cygnus X-3. From \citet{2009Sci...326.1512F}.}
\label{cygx3}
\end{center}
\end{figure}

A key recent result, however, was that the high-energy gamma-ray
emission from Cygnus X-3, as measured with \textit{Fermi}, was found
to be correlated with periods of radio flaring (observed with AMI-LA,
see Figure~\ref{cygx3}). This clearly established a direct link
between the presence of strong jet activity and the production of high
energy gamma-rays \citep{2009Sci...326.1512F}.

\subsubsection{GRS 1915+105}

In 1992, a new bright BH XRB, GRS 1915+105, was discovered. This
source became famous as the first source of apparently superluminal
jets in the Galaxy \citep{1994Natur.371...46M}, and has remained
active at close to its Eddington luminosity ever since.

The X-ray behaviour of GRS 1915+105 is extraordinary. The source
switches rapidly between three different X-ray `states', associated
with which are clearly correlated changes in the radio emission,
indicating that the jet can be launched, vary, and be shut off all on
timescales of minutes. The vast array of simultaneous X-ray, infrared
and radio observations of GRS 1915+105 are the most important single
resource, {\em in the whole of astrophysics}, for understanding the
connection between accretion and outflow.

\begin{figure}[t]
\begin{center}
\includegraphics[width=11.0cm]{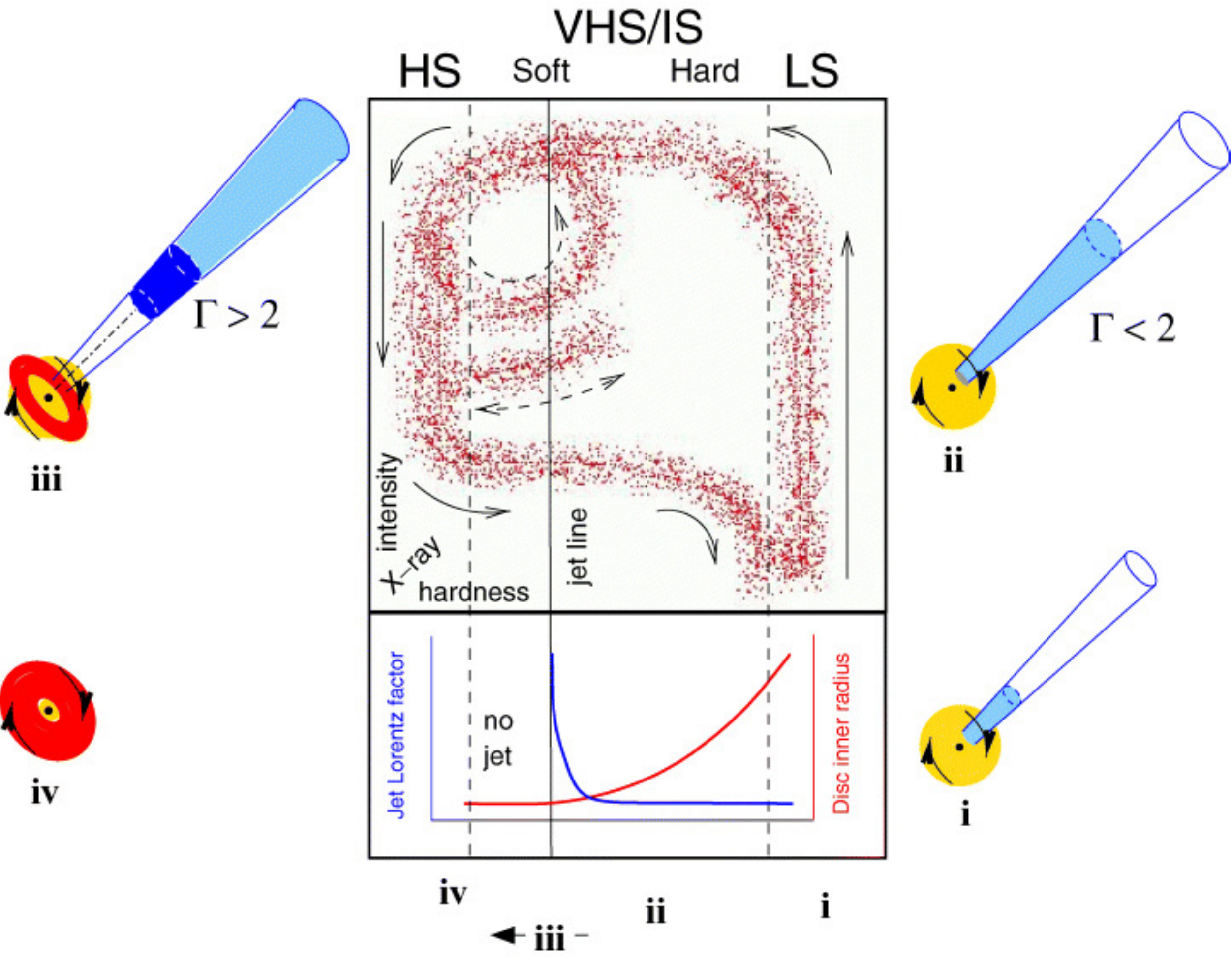} 
\caption{The unified empirical model for the coupling between
accretion and jet formation in BH XRBs from
\citet{2004MNRAS.355.1105F}. The model has been tested extensively and
seems to hold, and to also be applicable to AGN. Development of the
model was led in large part by the vast AMI-LA database on the
highly-variable BH GRS 1915+105.}
\label{1915+105}
\end{center}
\end{figure}

Careful study of these data led, eventually, to the unified empirical
model for the connection between accretion and jets in black holes
published in \citet{2004MNRAS.355.1105F} (see
Figure~\ref{1915+105}). This model remains consistent with all data to
the present day, and appears to be applicable in large part to
supermassive accreting black holes in AGN \citet{2006MNRAS.372.1366K}.

\subsubsection{Tidal-Disruption Event}

Many transient events, mostly X-ray binaries, SNe, AGN or GRB, have
been observed with the AMI-LA (and its predecessors). The source
\textit{Swift} J164449.3+573451, initially thought to be a GRB, is now
believed to be an exceptionally unusual tidal-disruption event: a star
that has been captured by an otherwise quiescent BH in a galactic
nucleus. The radio emission at 15.7 GHz has peaked after some 130
days, and is now declining gently. It should be visible for some
years, and may well be resolvable by VLBI instruments in due course.
There is a wide range of astrophysical information to be extracted
from the current set of observations, of which the AMI-LA is one of the
best sampled.

\subsection{Contribution of AMI in The Future}

AMI has demonstrated itself to be an extremely efficient, effective
and important radio telescope for the follow-up and monitoring of
high-energy astrophysical transients. This is for a number of reasons,
which include:

\begin{enumerate}
\item \label{future_1}\textbf{Highly-flexible scheduling} ---
  possibility of both rapid response and regular monitoring.
\item \label{future_2}\textbf{High radio frequency} --- astrophysical
events vary more rapidly, and more in tune with other wavelengths, at
higher radio frequencies. This is because the lower optical depth
probes closer to the jet `launch site'. The insights we made into the
disk-jet coupling in GRS 1915+105 and other sources would simply not
have been possible at, say, 1\,GHz, where the variations are `blurred'
out in time and width from the launch event. At 15.7\,GHz there is
also much less confusion and long runs are not required to measure
`core' radio-flux densities.
\end{enumerate}

Future operations of AMI will exploit and develop this capability.

Regarding~(\ref{future_1}), Cambridge and Southampton Universities
have recently developed software for the automatic triggering of an
AMI override observation based upon a VO Event alert (see e.g.~
skyalert.org and Section~\ref{ToO}).  This system has been used for
`target-of-opportunity' observations towards M31 ULX-1, GRB120305A,
GRB120308A and GRB120311A.  In parallel, development has begun on
software to automatically parse VO Events (xml format), make
evaluations and decisions and ultimately in some cases trigger an
observation. This work is part of a broader effort towards coordinated
and intelligent transient observations which includes teams from
LOFAR, MeerKAT, the EVN and the VO. Since AMI's scheduled observations
involve comparatively long, repeated measurements, such `targets of
opportunity' can generally be observed \emph{promptly}, with little
impact on existing programmes.

Regarding~(\ref{future_2}), although there will in some cases be merit
in triggering AMI based upon LOFAR, in most cases the exciting
astrophysical event will have happened significantly earlier than the
point at which the MHz radio emission peaks (specifically this is true
for synchrotron emission; some events may have very early coherent
bursts). The delay between emission peaks at GHz and MHz frequencies
can be weeks for XRBs, and years for GRB afterglows. Triggering on
{\em radio} (as opposed to e.g.~X-ray or gamma-ray) transients should
therefore be based on detections at as high a radio frequency as
possible. In this case APERTIF, the focal plane array upgrade to WSRT,
which will provide a $\sim 3$ deg$^{2}$ field of view in the 1-GHz
band, will probably be superior to LOFAR for providing transient
detections that are scientifically complemented by AMI data. APERTIF
has several proposed key science projects in the area of transients
(to be evaluated in 2012). Furthermore there is overlap of the sky
visible to AMI with MeerKAT and ASKAP, and the test arrays KAT-7 and
BETA. Both SKA precursors have embraced transient detection with both
image-plane and time-series approaches approved and under
development. Interestingly, the survey figures of merit for ASKAP,
APERTIF, MeerKAT and also the JVLA are all comparable, indicating
similar potential GHz transient yields (depending on observing
strategy) over the coming decade.  Unlike gamma-ray and X-ray
transients (with some exceptions, notably those found by \textit{Swift}),
which are typically localised (initially) only to a few degrees (and
therefore unsuitable for a simple rapid AMI follow-up without more
complex tiling), these radio transients will be localised with
$\simeq$arcsec accuracy.

\section{Upgrades to AMI}\label{upgrades}

\subsection{Automatic Target-of-Opportunity Facility}\label{ToO}

At the beginning of 2012, a collaboration between Southampton and
Cambridge Universities developed a fully automated rapid follow-up
facility for observation of transient events. As a first
implementation of this new ``robotic facility'' we have focused on
follow-up of the well localised GRB events observed by the \textit{Swift}
burst-alert telescope \citet{2005SSRv..120..143B}.  Staley et al. 2012
(in prep) will describe the AMI observations of the first ten \textit{Swift}
GRB targets, seven of which were initiated within minutes of the GRB
alert being distributed via the NASA GCN network.  This automatic
target-of-opportunity (ToO) facility comprises three subsystems:

\begin{itemize}

\item The distribution network, making use of the VOEvent standard
  \citep{2006ASPC..351..637W}  implemented through a Python software
  package (`Comet'; https://github.com/jdswinbank/Comet).
  We monitor new GRB events detected by the \textit{Swift} burst alert
  telescope, via a direct connection to the recently announced GCN
  VOEvents service. This approach can be easily extended as more sources
  of transient event data become available.

\item The observation request trigger, which uses a simple filter to
extract the required information and then generates a specially
formatted observation request email which is sent to the AMI control
computers.

\item The automated telescope response, which is implemented through
  an in-house program, {\sc rqcheck} that validates the email, checks
  syntax and keyword values, and then:
\begin{itemize}
 \item Checks if target is currently on-sky and is unobscured by
   the Sun or Moon.
 \item Determines the observation time according to the
   timing specified in the request (ASAP, next transit, or specified
   sidereal time).
  \item Finds the nearest phase calibrator from VLBA and JVAS catalogues.
  \item Checks the availability of the telescope, e.g. in case of
  priority observing or engineering work. 
  \item Constructs an entry for the observing queue.  
  \item Inserts this entry in the queue, rescheduling other entries.
 \item Notifies the requester of the outcome by e-mail.
\end{itemize}

\end{itemize}

Our current trigger policy is to issue an observation request to AMI-LA
for all \textit{Swift} GRB alerts with a declination above -10$^{\circ}$.  We
follow up GRBs which are detected during the ToO observation with an
ad-hoc schedule determined individually for each source.  For radio
non-detections we employ a logarithmic follow-up schedule with
additional observations after 1-2 days, 3-4 days, 1 week, 2 weeks, and
finally at 1 month.

\subsection{Digital Correlator Upgrade}

The current AMI correlator is an analogue lag, Fourier-transform
spectrometer, which synthesises eight 0.75-GHz channels from sixteen
sampled measurements of the real part of each baseline's correlation
at 25mm delays.  The 0.75-GHz wide channels are narrow enough to
prevent bandwidth smearing, but we find that they provide insufficient
spectral rejection to remove the effects of transmissions from
geostationary satellites, even with hardware filters in both the RF
and IF chains.  At present, transmissions from geostationary
satellites hamper AMI observations at declinations below 20 degrees:
at dec 15 degrees, about ten percent of time-domain data have to be
excised, and by 0 degrees 95 percent of data are corrupted. This is a
major problem because the equatorial sky is accessible to both
northern and southern telescopes and, by the same token, deep legacy
fields in many wavebands are equatorial. A digital correlator would
offer many more, much narrower channels over the observing band, and
spectral filtering would becomes feasible, improving the situation
dramatically. The technology of analogue-to-digital converters (ADCs)
has rapidly advanced to the extent that 4-bit, 10 giga-sample per
second ADCs are available at modest cost, making an affordable digital
correlator for AMI a possibility. As part of a successful ERC Starting
Grant application, Scaife (Southampton) has secured full funding to
allow the construction of a new digital correlator and this upgrade to
AMI will give AMI exciting new capabilities:

\begin{itemize}
\item Most importantly AMI will be able to image equatorial regions of
the sky which are currently inaccessible. This will allow many new
science programmes with observations towards key extragalactic deep
fields (e.g. COSMOS, Subaru/XMM-Newton Deep Survey UKIDSS-Ultra-Deep
Survey, VIDEO) and AMI will be able to image sky visible to both ALMA
and VISTA.
\item Spectral filtering of the visibilities will reduce
the fraction of data excised for RFI, increasing the frequency lever-arm 
for spectral work such as AME and increasing the overall sensitivity.
\item The new correlator will not suffer from systematics associated with 
lag spacing errors in the Fourier-transform spectrometer, and this will
increase the imaging dynamic range, allowing robust removal of much brighter
contaminating radio sources in SZ fields.  
\item The increased spectral resolution of the correlator will allow for
better identification of contamination and so reduce systematics in the 
final images.
\end{itemize}

\section{Future Operational Model}\label{future}

Until now the scientific direction for AMI has been determined by the
Cavendish Astrophysics Group in Cambridge, although a great many
external collaborations have been formed and have yielded the broad
variety of science described above.  The aim now is to change the
operational model for AMI, opening up the telescope to external
observers, to maximise new scientific return. A consortium of UK
universities will bid to the STFC for operations funding and will
determine the scientific programme for AMI, inviting observing proposals
from the community. This will allow new programmes to be initiated to
exploit the unique capabilities for low-surface-brightness,
centimetre-wavelength observation offered by AMI.


\section{References}

\begin{thebibliography}{99}

\bibitem[\protect\citeauthoryear{Acord, Churchwell, 
\& Wood}{1998}]{1998ApJ...495L.107A} Acord J.~M., Churchwell E., Wood
  D.~O.~S., 1998, ApJ, 495, L107  

\bibitem[\protect\citeauthoryear{Ali-Ha{\"i}moud, Hirata, 
\& Dickinson}{2009}]{2009MNRAS.395.1055A} Ali-Ha{\"i}moud Y., Hirata C.~M.,
  Dickinson C., 2009, MNRAS, 395, 1055  


\bibitem[\protect\citeauthoryear{Allen, Evrard, 
\& Mantz}{2011}]{2011ARA&A..49..409A} Allen S.~W., Evrard A.~E., Mantz A.~B.,
  2011, ARA\&A, 49, 409  

\bibitem[\protect\citeauthoryear{AMI Consortium: Ainsworth et 
al.}{2012}]{2012MNRAS.423.1089A} AMI Consortium: Ainsworth R. E., et
  al., 2012, MNRAS, 423,  1089 

\bibitem[\protect\citeauthoryear{AMI Consortium: Barker et 
al.}{2006}]{ami06} AMI Consortium: Barker R. et al.,
  2006, MNRAS, 369, L1 
 
\bibitem[\protect\citeauthoryear{AMI Consortium: Davies et
al.}{2011}]{2011MNRAS.415.2708A} AMI Consortium: Davies M. et al., 2011,
  MNRAS, 415, 2708 

\bibitem[\protect\citeauthoryear{AMI Consortium: Franzen et 
al.}{2011}]{2011MNRAS.415.2699A} AMI Consortium: Franzen T. et al.,
  2011, MNRAS, 415, 2699 

\bibitem[\protect\citeauthoryear{AMI Consortium: Hurley-Walker et
    al.}{2011}]{tash11} AMI Consortium: Hurley-Walker N.,
    et al., 2011, MNRAS, 414,  L75

\bibitem[\protect\citeauthoryear{AMI Consortium: Hurley-Walker et 
al.}{2012}]{tash12} AMI Consortium: Hurley-Walker N. et al.,
  2012, MNRAS, 419,  2921 

\bibitem[\protect\citeauthoryear{AMI Consortium: Olamaie et 
al.}{2012}]{2012MNRAS.421.1136A} AMI Consortium: Olamaie M. et al., 2012,
  MNRAS, 421, 1136 

\bibitem[\protect\citeauthoryear{AMI Consortium:
    Rodr\'{i}guez-Gonz\'{a}lvez C. et al.}{2011}]{carmen11a}
    AMI Consortium: Rodr\'{i}guez-Gonz\'{a}lvez C. et al., 2011,
    MNRAS, 414, 3751 

\bibitem[\protect\citeauthoryear{AMI Consortium:
    Rodr\'{i}guez-Gonz\'{a}lvez C. et al.}{2012}]{carmen11b}
    AMI Consortium: Rodr\'{i}guez-Gonz\'{a}lvez C. et al., 2012, MNRAS, 3459 

\bibitem[\protect\citeauthoryear{AMI Consortium: Scaife et
    al.}{2008}]{2008MNRAS.385..809S}  AMI Consortium: Scaife A.~M.~M., et al.,
  2008, MNRAS, 385, 809  

\bibitem[\protect\citeauthoryear{AMI Consortium: Scaife et
    al.}{2009}]{2009MNRAS.394L..46A} AMI Consortium: Scaife A.~M.~M., et al.,
  2009, MNRAS, 394, L46
 
\bibitem[\protect\citeauthoryear{AMI Consortium: Scaife et
    al.}{2009}]{2009MNRAS.400.1394A} AMI Consortium: Scaife A.~M.~M., et al.,
 2009, MNRAS, 400, 1394

\bibitem[\protect\citeauthoryear{AMI Consortium: Scaife et
    al.}{2010}]{2010MNRAS.403L..46S}  
AMI Consortium: Scaife A.~M.~M., et al., 2010, MNRAS, 403, L46 

\bibitem[\protect\citeauthoryear{AMI Consortium: Scaife et
    al.}{2010}]{2010MNRAS.406L..45S}  
AMI Consortium: Scaife A.~M.~M., et al., 2010, MNRAS, 406, L45 

\bibitem[\protect\citeauthoryear{AMI Consortium: Scaife et
    al.}{2011}]{2011MNRAS.410.2662A} AMI Consortium: Scaife A.~M.~M., et al.,
  2011, MNRAS, 410, 2662 

\bibitem[\protect\citeauthoryear{AMI Consortium: Scaife et
    al.}{2011}]{2011MNRAS.415..893A} AMI Consortium: Scaife A.~M.~M., et al.,
  2011, MNRAS, 415, 893 

\bibitem[\protect\citeauthoryear{Ami Consortium: Scaife et  
al.}{2012}]{2012MNRAS.420.1019A} AMI Consortium: Scaife A.~M.~M., et al., 
2012, MNRAS, 420, 1019 

 
\bibitem[\protect\citeauthoryear{AMI Consortium: Scaife et
    al.}{2011}]{2011arXiv1111.5184S}  AMI Consortium: Scaife A.~M.~M., et al.,
  2011, arXiv, arXiv:1111.5184  

\bibitem[\protect\citeauthoryear{AMI Consortium: Shimwell et 
al.}{2012}]{tim10} AMI Consortium: Shimwell T., et al.,
  2012, 2012, MNRAS, 423, 1463 

\bibitem[\protect\citeauthoryear{AMI Consortium: Zwart et
    al.}{2008}]{jon08}  
AMI Consortium: Zwart J.~T.~L., et al., 2008, MNRAS, 391, 1545

\bibitem[\protect\citeauthoryear{AMI Consortium: Zwart et
    al.}{2011}]{jon11}  
AMI Consortium: Zwart J.~T.~L., et al., 2011, MNRAS, 418, 2754 

\bibitem[\protect\citeauthoryear{Andre, Ward-Thompson 
\& Barsony}{1993}]{1993ApJ...406..122A} Andre P., Ward-Thompson D., Barsony
  M., 1993, ApJ, 406, 122  

\bibitem[\protect\citeauthoryear{Andrews 
\& Williams}{2005}]{2005ApJ...631.1134A} Andrews S.~M., Williams J.~P., 2005, ApJ, 631, 1134 

\bibitem[\protect\citeauthoryear{Anglada et 
al.}{1995}]{1995ApJ...443..682A} Anglada G., Estalella R., Mauersberger R., 
Torrelles J.~M., Rodriguez L.~F., Canto J., Ho P.~T.~P., D'Alessio P., 
1995, ApJ, 443, 682 


\bibitem[\protect\citeauthoryear{Arnaud et 
al.}{2010}]{2010A&A...517A..92A} Arnaud M., Pratt G.~W., Piffaretti
  R., B{\"o}hringer H., Croston J.~H., Pointecouteau E., 2010, A\&A,
  517, A92 

\bibitem[\protect\citeauthoryear{Barthelmy et 
al.}{2005}]{2005SSRv..120..143B} Barthelmy S.~D., et al., 2005, SSRv, 120, 
143 

\bibitem[\protect\citeauthoryear{Beckwith 
\& Sargent}{1993}]{1993ApJ...402..280B} Beckwith S.~V.~W., Sargent A.~I.,
  1993, ApJ, 402, 280  

\bibitem[\protect\citeauthoryear{Beltr{\'a}n et 
al.}{2011}]{2011A&A...532A..91B} Beltr{\'a}n M.~T., Cesaroni R., Zhang Q.,
  Galv{\'a}n-Madrid R., Beuther H., Fallscheer C., Neri R., Codella C., 2011,
  A\&A, 532, A91  

\bibitem[\protect\citeauthoryear{Beltr{\'a}n et 
al.}{2006}]{2006Natur.443..427B} Beltr{\'a}n M.~T., Cesaroni R., Codella 
C., Testi L., Furuya R.~S., Olmi L., 2006, Natur, 443, 427 


\bibitem[\protect\citeauthoryear{Bennett et 
al.}{2003}]{2003ApJS..148...97B} Bennett C.~L., et al., 2003, ApJS, 148, 97 

\bibitem[\protect\citeauthoryear{Berger et al.}{2012}]{2012ApJ...748...36B} 
Berger E., Zauderer A., Pooley G.~G., Soderberg A.~M., Sari R., Brunthaler 
A., Bietenholz M.~F., 2012, ApJ, 748, 36 

\bibitem[\protect\citeauthoryear{Bonnell et 
al.}{2011}]{2011MNRAS.410.2339B} Bonnell I.~A., Smith R.~J., Clark P.~C., 
Bate M.~R., 2011, MNRAS, 410, 2339 

\bibitem[\protect\citeauthoryear{Boss}{1998}]{boss1998} Boss 
A.~P., 1998, ApJ, 503, 923 

\bibitem[\protect\citeauthoryear{Bourke et al.}{2006}]{bourke2006} 
Bourke T.~L., et al., 2006, ApJ, 649, L37 


\bibitem[\protect\citeauthoryear{Bourke et al.}{2005}]{bourke2005} 
Bourke T.~L., Crapsi A., Myers P.~C., Evans N.~J., II, Wilner D.~J., Huard 
T.~L., J{\o}rgensen J.~K., Young C.~H., 2005, ApJ, 633, L129 

\bibitem[\protect\citeauthoryear{Clowe et al.}{2006}]{2006ApJ...648L.109C} 
Clowe D., Brada{\v c} M., Gonzalez A.~H., Markevitch M., Randall S.~W., 
Jones C., Zaritsky D., 2006, ApJ, 648, L109

\bibitem[\protect\citeauthoryear{Crapsi et 
al.}{2005}]{crapsi2005} Crapsi A., et al., 2005, A\&A, 439, 1023 

\bibitem[\protect\citeauthoryear{Davies}{2006}]{2006cmb..confE..18D} Davies 
R.~D., 2006, cmb..conf,  

\bibitem[\protect\citeauthoryear{Davis et al.}{2008}]{2008MNRAS.387..954D} 
Davis C.~J., Scholz P., Lucas P., Smith M.~D., Adamson A., 2008, MNRAS, 
387, 954 


\bibitem[\protect\citeauthoryear{Draine 
\& Lazarian}{1998}]{1998ApJ...494L..19D} Draine B.~T., Lazarian A., 1998, ApJ,
  494, L19 

\bibitem[\protect\citeauthoryear{Draine 
\& Lazarian}{1998}]{1998ApJ...508..157D} Draine B.~T., Lazarian A., 1998, ApJ,
  508, 157 
\bibitem[\protect\citeauthoryear{Dunham et al.}{2008}]{2008ApJS..179..249D} 
Dunham M.~M., Crapsi A., Evans N.~J., II, Bourke T.~L., Huard T.~L., Myers 
P.~C., Kauffmann J., 2008, ApJS, 179, 249 

\bibitem[\protect\citeauthoryear{Dutrey, Guilloteau, 
\& Simon}{2003}]{2003A&A...402.1003D} Dutrey A., Guilloteau S., Simon M.,
  2003, A\&A, 402, 1003  

\bibitem[\protect\citeauthoryear{Erickson}{1957}]{1957ApJ...126..480E} 
Erickson W.~C., 1957, ApJ, 126, 480 

\bibitem[\protect\citeauthoryear{Fender, Belloni, 
\& Gallo}{2004}]{2004MNRAS.355.1105F} Fender R.~P., Belloni T.~M.,
  Gallo E., 2004, MNRAS, 355, 1105  

\bibitem[\protect\citeauthoryear{Fermi LAT Collaboration et 
al.}{2009}]{2009Sci...326.1512F} Fermi LAT Collaboration, et al., 2009, 
Sci, 326, 1512 

\bibitem[\protect\citeauthoryear{Feroz et al.}{2009}]{2009MNRAS.398.2049F} 
Feroz F., Hobson M.~P., Zwart J.~T.~L., Saunders R.~D.~E., Grainge 
K.~J.~B., 2009, MNRAS, 398, 2049 

\bibitem[\protect\citeauthoryear{Franco-Hern{\'a}ndez 
\& Rodr{\'{\i}}guez}{2004}]{2004ApJ...604L.105F} Franco-Hern{\'a}ndez R.,
  Rodr{\'{\i}}guez L.~F., 2004, ApJ, 604, L105  


\bibitem[\protect\citeauthoryear{Gallo et al.}{2005}]{2005Natur.436..819G} 
Gallo E., Fender R., Kaiser C., Russell D., Morganti R., Oosterloo T., 
Heinz S., 2005, Natur, 436, 819 

\bibitem[\protect\citeauthoryear{Galv{\'a}n-Madrid et 
al.}{2009}]{2009ApJ...706.1036G} Galv{\'a}n-Madrid R., Keto E., Zhang Q., 
Kurtz S., Rodr{\'{\i}}guez L.~F., Ho P.~T.~P., 2009, ApJ, 706, 1036 

\bibitem[\protect\citeauthoryear{Galv{\'a}n-Madrid et 
al.}{2011}]{2011MNRAS.416.1033G} Galv{\'a}n-Madrid R., Peters T., Keto 
E.~R., Mac Low M.-M., Banerjee R., Klessen R.~S., 2011, MNRAS, 416, 1033 


\bibitem[\protect\citeauthoryear{Gou et al.}{2011}]{2011ApJ...742...85G} 
Gou L., et al., 2011, ApJ, 742, 85 

\bibitem[\protect\citeauthoryear{Greaves 
\& Rice}{2011}]{2011MNRAS.412L..88G} Greaves J.~S., Rice W.~K.~M., 2011,
  MNRAS, 412, L88  


\bibitem[\protect\citeauthoryear{Harvey et al.}{2007}]{2007ApJ...663.1149H} 
Harvey P., Mer{\'{\i}}n B., Huard T.~L., Rebull L.~M., Chapman N., Evans 
N.~J., II, Myers P.~C., 2007, ApJ, 663, 1149 


\bibitem[\protect\citeauthoryear{Hatchell 
\& Dunham}{2009}]{2009A&A...502..139H} Hatchell J., Dunham M.~M., 2009, A\&A,
  502, 139  

\bibitem[\protect\citeauthoryear{Kamp 
\& Dullemond}{2004}]{2004ApJ...615..991K} Kamp I., Dullemond C.~P., 2004, ApJ,
  615, 991  

\bibitem[\protect\citeauthoryear{Kenyon 
\& Hartmann}{1990}]{1990ApJ...349..197K} Kenyon S.~J., Hartmann L.~W., 1990,
  ApJ, 349, 197  

\bibitem[\protect\citeauthoryear{Klaassen et 
al.}{2009}]{2009ApJ...703.1308K} Klaassen P.~D., Wilson C.~D., Keto E.~R., 
Zhang Q., 2009, ApJ, 703, 1308 

\bibitem[\protect\citeauthoryear{Kogut et al.}{1996}]{1996ApJ...464L...5K} 
Kogut A., Banday A.~J., Bennett C.~L., Gorski K.~M., Hinshaw G., Smoot 
G.~F., Wright E.~I., 1996, ApJ, 464, L5 
 
\bibitem[\protect\citeauthoryear{Komatsu et 
al.}{2011}]{2011ApJS..192...18K} Komatsu E., et al., 2011, ApJS, 192,
  18 

\bibitem[\protect\citeauthoryear{K{\"o}rding, Jester, 
\& Fender}{2006}]{2006MNRAS.372.1366K} K{\"o}rding E.~G., Jester S.,
  Fender R., 2006, MNRAS, 372, 1366  

\bibitem[\protect\citeauthoryear{Korngut et 
al.}{2011}]{2011ApJ...734...10K} Korngut P.~M., et al., 2011, ApJ, 734, 10 

\bibitem[\protect\citeauthoryear{Kravtsov, Vikhlinin, 
\& Nagai}{2006}]{2006ApJ...650..128K} Kravtsov A.~V., Vikhlinin A., Nagai D.,
  2006, ApJ, 650, 128 

\bibitem[\protect\citeauthoryear{Kuiper et al.}{2011}]{2011ApJ...732...20K} 
Kuiper R., Klahr H., Beuther H., Henning T., 2011, ApJ, 732, 20 

\bibitem[\protect\citeauthoryear{Kurtz}{2005}]{2005IAUS..227..111K} Kurtz 
S., 2005, IAUS, 227, 111 

\bibitem[\protect\citeauthoryear{Kunz et al.}{2011}]{2011MNRAS.410.2446K} 
Kunz M.~W., Schekochihin A.~A., Cowley S.~C., Binney J.~J., Sanders J.~S., 
2011, MNRAS, 410, 2446 

\bibitem[\protect\citeauthoryear{Leitch et al.}{1997}]{1997ApJ...486L..23L} 
Leitch E.~M., Readhead A.~C.~S., Pearson T.~J., Myers S.~T., 1997, ApJ, 
486, L23 

\bibitem[\protect\citeauthoryear{Longmore et 
al.}{2007}]{2007MNRAS.379..535L} Longmore S.~N., Burton M.~G., Barnes 
P.~J., Wong T., Purcell C.~R., Ott J., 2007, MNRAS, 379, 535 

\bibitem[\protect\citeauthoryear{Lueker et al.}{2010}]{2010ApJ...719.1045L} 
Lueker M., et al., 2010, ApJ, 719, 1045 

\bibitem[\protect\citeauthoryear{Marriage et 
al.}{2011}]{2011ApJ...737...61M} Marriage T.~A., et al., 2011, ApJ, 737, 61 


\bibitem[\protect\citeauthoryear{Mason et al.}{2010}]{2010ApJ...716..739M} 
Mason B.~S., et al., 2010, ApJ, 716, 739 

\bibitem[\protect\citeauthoryear{McKee 
\& Tan}{2003}]{2003ApJ...585..850M} McKee C.~F., Tan J.~C., 2003, ApJ,
585, 850

\bibitem[\protect\citeauthoryear{Mirabel 
\& Rodr{\'{\i}}guez}{1994}]{1994Natur.371...46M} Mirabel I.~F.,
  Rodr{\'{\i}}guez L.~F., 1994, Natur, 371, 46  

\bibitem[\protect\citeauthoryear{Molinari et 
al.}{2010}]{2010A&A...518L.100M} Molinari S., et al., 2010, A\&A, 518, L100 

\bibitem[\protect\citeauthoryear{Peters et al.}{2010}]{2010ApJ...719..831P} 
Peters T., Mac Low M.-M., Banerjee R., Klessen R.~S., Dullemond C.~P., 
2010, ApJ, 719, 831

\bibitem[\protect\citeauthoryear{Planck Collaboration: Ade et 
al.}{2011}]{2011A&A...536A...8P} Planck Collaboration: Ade P. A. R., et
  al., 2011, A\&A, 536, A8  

\bibitem[\protect\citeauthoryear{Planck Collaboration: Ade et
al.}{2011}]{2011A&A...536A..20P} Planck Collaboration: Ade P. A. R., et al.,
  2011, A\&A, 536, A20 

\bibitem[\protect\citeauthoryear{Planck Collaboration: Aghanim et 
al.}{2011}]{2011A&A...536A..10P} Planck Collaboration: Aghanim N., et
  al., 2011, A\&A, 536, A10  

\bibitem[\protect\citeauthoryear{Planck Collaboration: Aghanim et 
al.}{2012}]{arXiv:1204.1318} Planck Collaboration: Aghanim N., et
  al., 2012, arXiv:1204.1318  

\bibitem[\protect\citeauthoryear{Pooley, Fender, 
\& Brocksopp}{1999}]{1999MNRAS.302L...1P} Pooley G.~G., Fender R.~P.,
  Brocksopp C., 1999, MNRAS, 302, L1  

\bibitem[\protect\citeauthoryear{Purcell 
\& Hoare}{2010}]{2010HiA....15..781P} Purcell C.~R., Hoare M.~G., 2010, HiA,
  15, 781  

\bibitem[\protect\citeauthoryear{Reichardt et 
al.}{2012}]{2012arXiv1203.5775R} Reichardt C.~L., et al., 2012, arXiv, 
arXiv:1203.5775 

\bibitem[\protect\citeauthoryear{Shirley et 
al.}{2007}]{2007ApJ...667..329S} Shirley Y.~L., Claussen M.~J., Bourke 
T.~L., Young C.~H., Blake G.~A., 2007, ApJ, 667, 329 

\bibitem[\protect\citeauthoryear{Shirley et 
al.}{2011}]{2011ApJ...728..143S} Shirley Y.~L., Huard T.~L., Pontoppidan 
K.~M., Wilner D.~J., Stutz A.~M., Bieging J.~H., Evans N.~J., II, 2011, 
ApJ, 728, 143 

\bibitem[\protect\citeauthoryear{Shirley et 
al.}{2011}]{2011AJ....141...39S} Shirley Y.~L., Mason B.~S., Mangum J.~G., 
Bolin D.~E., Devlin M.~J., Dicker S.~R., Korngut P.~M., 2011, AJ, 141, 39 

\bibitem[\protect\citeauthoryear{Sievers et 
al.}{2009}]{2009arXiv0901.4540S} Sievers J.~L., et al., 2009, arXiv, 
arXiv:0901.4540 

\bibitem[\protect\citeauthoryear{Walawender, Bally, 
\& Reipurth}{2005}]{2005AJ....129.2308W} Walawender J., Bally J., Reipurth B.,
  2005, AJ, 129, 2308  

\bibitem[\protect\citeauthoryear{Williams 
\& Seaman}{2006}]{2006ASPC..351..637W} Williams R.~D., Seaman R., 2006, ASPC, 351, 637 

\bibitem[\protect\citeauthoryear{Zauderer et 
al.}{2011}]{2011Natur.476..425Z} Zauderer B.~A., et al., 2011, Natur, 476, 
425 

\bibitem[\protect\citeauthoryear{Zdziarski et 
al.}{2011}]{2011MNRAS.416.1324Z} Zdziarski A.~A., Skinner G.~K., Pooley 
G.~G., Lubi{\'n}ski P., 2011, MNRAS, 416, 1324 

\end{thebibliography}
\end{document}